\RequirePackage{etex}

\documentclass[pdftex,colorlinks,notitlepage,a4paper,12pt]{article}

\usepackage{slantsc}

\usepackage[a4paper]{geometry}

\usepackage[pdftex]{graphicx}

\usepackage{epstopdf}

\usepackage[usenames,dvipsnames]{color}
\usepackage[dvipsnames]{xcolor}

\usepackage{soul}

\newlength{\mylength}
\makeatletter
\newcommand{\mycfs}[1]{%
\normalsize
\@defaultunits\mylength=#1pt\relax\@nnil
\edef\@tempa{{\strip@pt\mylength}}%
\ifx\protect\@typeset@protect
\edef\@currsize{\noexpand\mycfs\@tempa}
\fi
\mylength=1.2\mylength
\edef\@tempa{\@tempa{\strip@pt\mylength}}%
\expandafter\fontsize\@tempa
\selectfont
}

\usepackage[miktex]{pdftricks2}

\usepackage[utf8]{inputenc}
\usepackage[T1]{fontenc}

\usepackage[french, german, spanish, british]{babel}
\babeltags{es = spanish}
\babeltags{de = german}
\usepackage[style=american]{csquotes}

\usepackage[en-GB]{datetime2}
\DTMlangsetup[en-GB]{ord=raise}

\usepackage{textgreek}

\usepackage{latexsym}

\usepackage{amsmath}

\usepackage{econometrics}

\usepackage{amssymb}

\usepackage{pdflscape} 

\usepackage{afterpage}

\usepackage{array}
\usepackage{delarray}
\usepackage{cases}
\usepackage{rotating}
\usepackage{longtable}
\usepackage{threeparttable}
\usepackage{threeparttablex}
\usepackage{booktabs}
\usepackage{caption}
\usepackage{subcaption}
\usepackage{makecell}

\usepackage{tabularx}
\usepackage{ltablex}
\keepXColumns
\usepackage{tabu}
\usepackage{linegoal}
\usepackage{multirow}

\makeatletter
\newlength\LongtableWidth
\newcommand*{\org@longtable}{}
\let\org@longtable\longtable
\def\longtable{%
\begingroup
\advance\c@LT@tables\@ne
\edef\x{LT@\romannumeral\c@LT@tables}
\global\LongtableWidth\z@
\@ifundefined{\x}{%
}{%
\def\LT@entry##1##2{%
\global\advance\LongtableWidth##2\relax
}%
\@nameuse{\x}%
}%
\typeout{* \x: \the\LongtableWidth}%
\endgroup \ifdim\LongtableWidth>\z@ \setlength{\LTcapwidth}{\LongtableWidth}%
\fi
\org@longtable }
\makeatother

\usepackage{ragged2e}

\usepackage{xpatch}

\usepackage{float}
\usepackage{placeins}

\usepackage{eurosym} 

\usepackage{appendix}

\usepackage[hyphens]{url}

\usepackage{xurl}

\hyphenchar\font=\string"7F
\usepackage[super]{nth}

\usepackage[shortlabels]{enumitem} 
\setlist[enumerate]{nosep, label = {(\arabic*)}}
\setlist[itemize]{nosep, label = {---}}

\usepackage{titlesec}
\titlelabel{\thetitle.\quad}
\titleformat*{\section}{\normalfont\large\bfseries\setstretch{1.5}}
\titleformat*{\subsection}{\normalfont\large\bfseries\setstretch{1}}

\usepackage{fancyhdr}

\makeatletter
\newread\pin@file
\newcounter{pinlineno}
\newcommand\pin@accu{}
\newcommand\pin@ext{pintmp}
\newcommand*\partialinput [3] {%
\IfFileExists{#3}{%
\openin\pin@file #3
\setcounter{pinlineno}{1}
\@whilenum\value{pinlineno}<#1 \do{%
\read\pin@file to\pin@line
\stepcounter{pinlineno}%
}
\addtocounter{pinlineno}{-1}
\let\pin@accu\empty
\begingroup
\endlinechar\newlinechar
\@whilenum\value{pinlineno}<#2 \do{%
\readline\pin@file to\pin@line
\edef\pin@accu{\pin@accu\pin@line}%
\stepcounter{pinlineno}%
}
\closein\pin@file
\expandafter\endgroup
\scantokens\expandafter{\pin@accu}%
}{%
\errmessage{File `#3' doesn't exist!}%
}%
}
\makeatother

\usepackage[backend=biber,style=apa,uniquelist=false,uniquename=false,dashed=true,dateabbrev=false,eventdate=comp,hyperref=true,language=british,parentracker=true]{biblatex}

\DeclareNameFormat{labelname:poss}{
\nameparts{#1}
\ifcase\value{uniquename}%
\usebibmacro{name:family}{\namepartfamily}{\namepartgiven}{\namepartprefix}{\namepartsuffix}%
\or
\ifuseprefix
{\usebibmacro{name:first-last}{\namepartfamily}{\namepartgiveni}{\namepartprefix}{\namepartsuffixi}}
{\usebibmacro{name:first-last}{\namepartfamily}{\namepartgiveni}{\namepartprefixi}{\namepartsuffixi}}%
\or
\usebibmacro{name:first-last}{\namepartfamily}{\namepartgiven}{\namepartprefix}{\namepartsuffix}%
\fi
\usebibmacro{name:andothers}%
\ifnumequal{\value{listcount}}{\value{liststop}}{'s}{}}
\DeclareFieldFormat{shorthand:poss}{%
\ifnameundef{labelname}{#1's}{#1}}
\DeclareFieldFormat{citetitle:poss}{\mkbibemph{#1}'s}
\DeclareFieldFormat{label:poss}{#1's}
\newrobustcmd*{\posscitealias}{%
\AtNextCite{%
\DeclareNameAlias{labelname}{labelname:poss}%
\DeclareFieldAlias{shorthand}{shorthand:poss}%
\DeclareFieldAlias{citetitle}{citetitle:poss}%
\DeclareFieldAlias{label}{label:poss}}}
\newrobustcmd*{\posscite}{%
\posscitealias%
\textcite}
\newrobustcmd*{\Posscite}{\bibsentence\posscite}
\newrobustcmd*{\posscites}{%
\posscitealias%
\textcites}

\DeclareFieldFormat*{edition}{\nth{#1} ed.}

\newcommand{\noop}[1]{}

\usepackage{siunitx} 

\def\yyy{%
\bgroup\uccode`\~\expandafter`\string-%
\uppercase{\egroup\edef~{\noexpand\text{\llap{\textendash}\relax}}}%
\mathcode\expandafter`\string-"8000 }

\def\xxxl#1{%
\bgroup\uccode`\~\expandafter`\string#1%
\uppercase{\egroup\edef~{\noexpand\text{\noexpand\llap{\string#1}}}}%
\mathcode\expandafter`\string#1"8000 }

\def\xxxr#1{%
\bgroup\uccode`\~\expandafter`\string#1%
\uppercase{\egroup\edef~{\noexpand\text{\noexpand\rlap{\string#1}}}}%
\mathcode\expandafter`\string#1"8000 }

\usepackage{setspace}
\usepackage[bottom]{footmisc}

\usepackage[framemethod=tikz]{mdframed}
\usepackage{tikz}
\usetikzlibrary{shapes.geometric, arrows, positioning, calc, decorations.markings, shapes.misc, fit}

\usepackage{acronym}
\acrodef{who}[WHO]{World Health Organization}
\acrodef{eclac}[ECLAC]{Economic Commission for Latin America and the Caribbean}
\acrodef{pis}[PIS]{perinatal information system}
\acrodef{did}[DID]{differences-in-differences}
\acrodef{oecd}[OECD]{Organisation for Economic Cooperation and Development}
\acrodef{ine}[INE]{Instituto Nacional de Estad{\'i}stica}
\acrodef{un}[UN]{United Nations}

\usepackage{etoolbox}
\makeatletter
\patchcmd{\maketitle}{\@makefntext}{\fakecommand}{}{}
\patchcmd{\maketitle}{\rlap}{\hbox}{}{}
\patchcmd{\@maketitle}{\@author}{\hspace*{5pt}\@author}{}{}
\makeatother

\usepackage{comment}

\usepackage{changes}

\definechangesauthor[color=Green]{NA}

\usepackage{todonotes}

\usepackage{abstract}
\usepackage{titling}

\usepackage[noblocks]{authblk}

\title{The short- and long-term determinants of fertility in Uruguay\thanks{Corresponding author: Jos{\'e}-Ignacio Ant{\'o}n, Department of Applied Economics, University of Salamanca, Campus Miguel de Unamuno, 37007 Salamanca (Spain), e-mail: \href{mailto:janton@usal.es}{\texttt{janton@usal.es}}. We thank Nicol{\'a}s Bonino, Fernando Borr{\'a}z, Elizabeth Bucacos and Wanda Cabella for their comments on a previous version of this paper. The authors acknowledge the financial support of the Spanish Ministry of Science and Innovation (project PID2021-123875NB-I00).}}

\pretitle{\vspace{-40pt}\centering\Large\bfseries}
\posttitle{\par\centering}

\predate{\centering}
\postdate{\centering\vskip -2em}

\makeatletter
\AtBeginDocument{%
\newlength{\temp@x}%
\newlength{\temp@y}%
\newlength{\temp@w}%
\newlength{\temp@h}%
\def\my@coords#1#2#3#4{%
\setlength{\temp@x}{#1}%
\setlength{\temp@y}{#2}%
\setlength{\temp@w}{#3}%
\setlength{\temp@h}{#4}%
\adjustlengths{}%
\my@pdfliteral{\strip@pt\temp@x\space\strip@pt\temp@y\space\strip@pt\temp@w\space\strip@pt\temp@h\space re}}%
\ifpdf
\typeout{In PDF mode}%
\def\my@pdfliteral#1{\pdfliteral page{#1}}
\def\adjustlengths{}%
\fi
\ifxetex
\def\my@pdfliteral #1{}
\def\adjustlengths{\setlength{\temp@h}{-\temp@h}\addtolength{\temp@y}{1in}\addtolength{\temp@x}{-1in}}%
\fi%
\def\Hy@colorlink#1{%
\begingroup
\ifHy@ocgcolorlinks
\def\Hy@ocgcolor{#1}%
\my@pdfliteral{q}%
\my@pdfliteral{7 Tr}
\else
\HyColor@UseColor#1%
\fi
}%
\def\Hy@endcolorlink{%
\ifHy@ocgcolorlinks%
\my@pdfliteral{/OC/OCPrint BDC}%
\my@coords{0pt}{0pt}{\pdfpagewidth}{\pdfpageheight}%
\my@pdfliteral{F}
\my@pdfliteral{EMC/OC/OCView BDC}%
\begingroup%
\expandafter\HyColor@UseColor\Hy@ocgcolor%
\my@coords{0pt}{0pt}{\pdfpagewidth}{\pdfpageheight}%
\my@pdfliteral{F}
\endgroup%
\my@pdfliteral{EMC}%
\my@pdfliteral{0 Tr}
\my@pdfliteral{Q}%
\fi
\endgroup
}%
}
\makeatother

\DeclareLabeldate{%
\field{date}
\field{eventdate}
\field{origdate}
\field{urldate}
\field{pubstate}
\literal{nodate}
}

\renewbibmacro*{addendum+pubstate}{%
\printfield{addendum}%
\iffieldequalstr{labeldatesource}{pubstate}{}
{\newunit\newblock\printfield{pubstate}}}

\author[$\dag$]{Zuleika~Ferre}
\affil[$\dag$]{University of the Republic}
\author[$\dag$]{Patricia~Triunfo}
\author[$\ddag$]{Jos\'e-Ignacio~Antón}
\affil[$\ddag$]{University of Salamanca (Spain) and Instituto Universitario Guti{\'e}rrez Mellado, Universidad Nacional de Educaci{\'o}n a Distancia (Spain)}

\usepackage[pdftex,pdftitle={},pdfsubject={},%
pdfauthor={},breaklinks=true]{hyperref}
\hypersetup{
colorlinks=true,%
citecolor=blue,%
filecolor=black,%
linkcolor=black,%
urlcolor=blue}

\begin{document}
\date{}

\maketitle

\singlespacing 


\begin{abstract}
\noindent This paper examines the determinants of fertility among women at different stages of their reproductive lives in Uruguay. To this end, we employ time series analysis methods based on data from 1968 to 2021 and panel data techniques based on department-level statistical information from 1984 to 2019. The results of our first econometric exercise indicate a cointegration relationship between fertility and economic performance, education and infant mortality, with differences observed by reproductive stage. We find a negative relationship between income and fertility for women aged 20--29 that persists for women aged 30 and over. This result suggests that having children is perceived as an opportunity cost for women in this age group. We also observe a negative relationship between education and adolescent fertility, which has implications for the design of public policies. A panel data analysis with econometric techniques allowing us to control for unobserved heterogeneity confirms that income is a relevant factor for all groups of women and reinforces the crucial role of education in reducing teenage fertility. We also identify a negative correlation between fertility and employment rates for women aged 30 and above. We outline some possible explanations for these findings in the context of work--life balance issues and argue for the importance of implementing social policies to address them.\vskip 0.5em

\end{abstract}

\singlespacing 

\section{Introduction}\label{Section 1}

Uruguay's fertility behaviour has idiosyncratic features. The country was one of the pioneers of the demographic transition in Latin America and the Caribbean, with very early declines in both fertility and mortality. Its fertility rate was 2.7 children per woman in 1950, a figure that the continent did not reach until the end of the \nth{20} century. Nevertheless, adolescent fertility---with its well-known negative public health and socioeconomic consequences---remained high until recently.\footnote{Adolescent fertility is particularly relevant because of its impact throughout the life of teenage mothers: it is due to low educational attainment, poor labour market outcomes and poverty \parencite{engelhardt2004b,fletcher2009,hoffman2008,lopez2016,paranjothy2009,varela2004,varela2014a,varela2014b}. Teenage pregnancies are often unplanned \parencite{anton2018,buckles2019}, receive less prenatal care and have worse birth outcomes on average \parencite{joyce1990,kost2015,kost2018,moreira2019}.} It peaked in 1997 at 74 births per thousand women, stabilised at approximately 60 births per thousand women in the following years and experienced a marked decline from 2014 to 2021, when it reached 26 births per thousand women \parencite{un2022}.

The literature has attempted to explain the determinants of fertility at different stages of reproductive life using different conceptual frameworks, methodological tools and types of data. The results are ambiguous, possibly influenced by the different approaches mentioned above, which underlines the importance of providing new empirical evidence that sheds light on this phenomenon. Only a proper understanding of the factors driving fertility can allow policy makers to implement policies that either encourage or discourage fertility, depending on the context. While the adolescent birth rate was high until recently, the total fertility rate reached its minimum in 2020 at 1.48 births per woman, well below replacement level \parencite{wb2023}. As a result, Uruguay is experiencing a rapid ageing process and, given its relatively high level of social protection for its per capita income, is inherently facing relevant fiscal pressures in the coming decades \parencite{rofman2016}. 

The aim of this paper is to investigate the determinants of fertility among women aged 15--19 (teenage fertility), 20--29 (intermediate fertility) and 30 and over (late fertility). We use both time series analysis methods, based on data from 1968 to 2021, and panel data techniques, based on department-level statistical information from 1984 to 2019. The results of our time series econometric exercise show the existence of a cointegration relationship between fertility and economic performance (GDP per capita and female employment), education and infant mortality. Our findings indicate that income levels have a negative impact on fertility rates at various stages of the reproductive cycle. Education, on the other hand, appears to be an effective means of reducing teenage fertility rates. In terms of employment, our study highlights its negative correlation with fertility rates among women aged 30 and over, underscoring the significance of the opportunity cost of motherhood. We offer some explanations for these findings from the perspective of work--life balance problems and social policies to address them.

Overall, our research emphasizes the importance of considering multiple factors in exploring fertility rates and highlights the need for effective policies that support family planning and work--life balance.

The literature includes examples of in-depth studies covering several decades using the types of techniques employed here and specifically focused on the United States \parencite{kearney2015b}, Japan \parencite{kato2021,suzuki2019} and Italy \parencite{cazzola2016}. Nevertheless, to the best our knowledge, there are no studies specifically on either Uruguay or any other Latin American or Caribbean country. In addition, this work aims to contribute to the existing literature on the subject of fertility behaviour by providing additional evidence that cumulatively helps to increase our knowledge of its determinants. 

After this introduction, the rest of the chapter unfolds as follows. The second section summarises the theoretical framework for studying fertility dynamics and explains the contributions of our work to the existing literature. The third and fourth sections cover the country and regional analyses, respectively, including a description of the methodology and data and presenting our results. Last, we discuss the main conclusions and implications of our study. 

\section{Theoretical framework and literature review}\label{Section 2}

Studies of the determinants of fertility over reproductive life make use of different conceptual frameworks and operate at at least four different levels of analysis (national, community, household and individual). We summarise and systematise these approaches in Figure~\ref{Figure 1}, drawing on the seminal contribution of \textcite{davis1956} and the later adaptations by \textcite{bongaarts1978} and \textcite{ojakaa2022}.

The model of \textcite{davis1956} proposes a set of intermediate determinants of fertility, inspiring the simplified approach of \textcite{bongaarts1978} that focuses on so-called proximate determinants of fertility. These factors differ according to the stage of reproductive life and comprise three categories: exposure (nuptiality and age of sexual initiation), deliberate control of fertility (access to and use of birth-control methods and abortion) and natural fertility. The latter factor refers to the absence of contraception and depends on women's exposure and reproductive conditions (such as sexual abstinence, age at first sexual intercourse, coital frequency, miscarriages, infertility and breastfeeding). In turn, these causes and dynamics of fertility imply several empirically testable hypotheses. 

\tikzset{->-/.style={decoration={markings, mark=at position #1 with {\arrow{>}}},postaction={decorate}}}
\tikzset{>=latex}
\begin{figure}[!htbp]
\caption{Determinants of fertility}
\vspace{-0.5em}
\begin{center}
\begin{tikzpicture}[decoration={markings, mark= at position 0.5 with {\arrow{>}}}] 
\scriptsize
\tikzstyle{node} = [rectangle, minimum width=\textwidth, minimum height=1cm]
\tikzstyle{line}=[draw, thick, -latex']
\node (1) [draw, align=left] (word1) [node] {\hskip 4.5cm\textbf{National-level determinants}\\
$\circ$ Health policies and programmes: family planning, abortion, sexual and reproductive health\\$\circ$ Other sectorial policies and programmes: education and employment\\$\circ$ Institutions};
\tikzstyle{node} = [rectangle,minimum width=0.45\textwidth , minimum height=1cm]
\node (2) [draw, align=left, text width = 0.4\textwidth] at (-4.04,-3.2) (word2) [node] {\textbf{Contextual-level variables (urban/rural)}\\$\circ$ Social and cultural norms\\$\circ$ Institutions\\$\circ$ Economic and environmental conditions\\$\circ$ Community socioeconomic characteristics (female labour market participation, female educational attainment, average age at first union)}; 
\node (3) [draw, align=left, text width = 0.4\textwidth] at (4.05,-2.5) (word3) [node] {\hskip 1cm\textbf{Household-level variables}\\$\circ$ Socioeconomic level\\$\circ$ Family structure (female household head, type of household)\\$\circ$ Religiosity\\$\circ$ Intergenerational relationships};
\tikzstyle{node} = [rectangle,minimum width=0.35\textwidth , minimum height=1cm] 
\node (4) [draw, align=left, text width=0.3\textwidth] at (4.77,-5.8) (word4) [node] {\hskip 0.2cm\textbf{Individual-level variables}\\$\circ$ Age\\$\circ$ Education\\$\circ$ Marital status\\$\circ$ Employment\\$\circ$ Religiosity\\$\circ$ Media exposure\\$\circ$ Knowledge of contraceptive methods}; 
\tikzstyle{node} = [rectangle,minimum width=0.42\textwidth, minimum height=1cm] 
\node (5) [draw, align=left, text width = 0.35\textwidth] at (-3.82,-9.5) (word5) [node] {\hskip 1.1cm\textbf{Fertility demand}\\$\circ$ Perceptions of desired number of children\\$\circ$ Preferences over and constraints on children\\$\circ$ Cost of access to and use of contraceptive methods}; 
\tikzstyle{node} = [rectangle,minimum width=0.42\textwidth , minimum height=2.64cm] 
\node (6) [draw, align=left, text width=0.35\textwidth] at (3.812,-9.5) (word6) [node] {\hskip 1.1cm\textbf{Fertility supply}\\[1.4ex]Ability to control fertility/natural fertility: infertility, interbirth interval, time to pregnancy, intrauterine mortality};
\tikzstyle{node} = [rectangle,minimum width=0.3\textwidth] 
\node (7) [draw, align=left, text width = 0.25\textwidth] at (-4.7,-14) (word7) [node] {\hskip 0.85cm\textbf{Teen fertility}\\$\circ$ Age at first sexual intercourse\\$\circ$ Age at first union\\$\circ$ Access to and use of contraceptive methods\\$\circ$ Natural fertility\\$\circ$ Infant mortality};
\tikzstyle{node} = [rectangle,minimum width=0.3\textwidth, minimum height=2.95cm]
\node (8) [draw, align=left, text width = 0.25\textwidth, right=0.27cm of word7] (word8) [node] {\hskip 0.35cm\textbf{Intermediate fertility}\\[0.9ex]$\circ$ Interbirth interval\\$\circ$ Access to and use of contraceptive methods\\$\circ$ Natural fertility\\$\circ$ Infant mortality};
\node (9) [draw, align=left, text width = 0.25\textwidth, right=0.27cm of word8] (word9) [node] {\hskip 1cm\textbf{Late fertility}\\[1.4ex]$\circ$ Access to and use of contraceptive methods\\$\circ$ Natural fertility\\$\circ$ Infant mortality};
\tikzstyle{node} = [rectangle, minimum width=\textwidth, minimum height=1cm]
\node (10) [draw, align=center, below=15.5cm of word1] (word10) [node] {Fertility};
\tikzstyle{node} = [rectangle, minimum width=\textwidth, minimum height=3.5cm]
\node (11) [draw=none, rounded rectangle, align=center, above=-1.4cm of word8] (word11) [node] {Proximate determinants by stage of reproductive life};
\draw [draw=black, rounded corners, dashed] (-7.345,-8.1) rectangle +(\textwidth, -2.8cm);
\draw [draw=black, rounded corners, dashed] (-7.345,-12.4) rectangle +(\textwidth, -3.2cm);
\draw [-latex,thick] ([xshift=-4.05cm]word1.south) -- (word2);
\draw [-latex,thick] ([xshift=4.05cm]word1.south) -- (word3);
\draw [-latex,thick] ([yshift=0.7cm]word2.east) -- (word3.west);
\draw [-latex,thick] (word3.south) -- ([xshift=-0.72cm]word4.north);
\draw [-latex,thick] ([xshift=-2.735cm]word3.south) -- ([xshift=-2.5cm,yshift=0.1cm]word6.north);
\draw [-latex,thick] ([xshift=-0.72cm]word4.south) -- ([yshift=0.1cm,xshift=0.24cm]word6.north);
\draw [-latex,thick] (word2.south) -- ([yshift=0.1cm,xshift=-0.22cm]word5.north); 
\draw [-latex,thick] ([yshift=-0.1cm,xshift=3.81cm]word5.south) -- ([yshift=0.7cm]word8.north); 
\draw [-latex,thick] ([yshift=-0.1cm]word7.south) -- ([xshift=-4.70cm]word10.north);
\draw [-latex,thick] ([yshift=-0.1cm,xshift=0.006cm]word8.south) -- (word10.north); 
\draw [-latex,thick] ([yshift=-0.1cm]word9.south) -- ([xshift=4.682cm]word10.north); 
\draw [latex-latex,thick] (word5.east) -- (word6.west); 
\end{tikzpicture}
\end{center}
\vspace{-0.5em}
\justifying
\footnotesize
\noindent\textit{Source}: Authors' elaboration from \textcite{bongaarts1978}, \textcite{davis1956} and \textcite{ojakaa2022}. 
\label{Figure 1}
\end{figure}
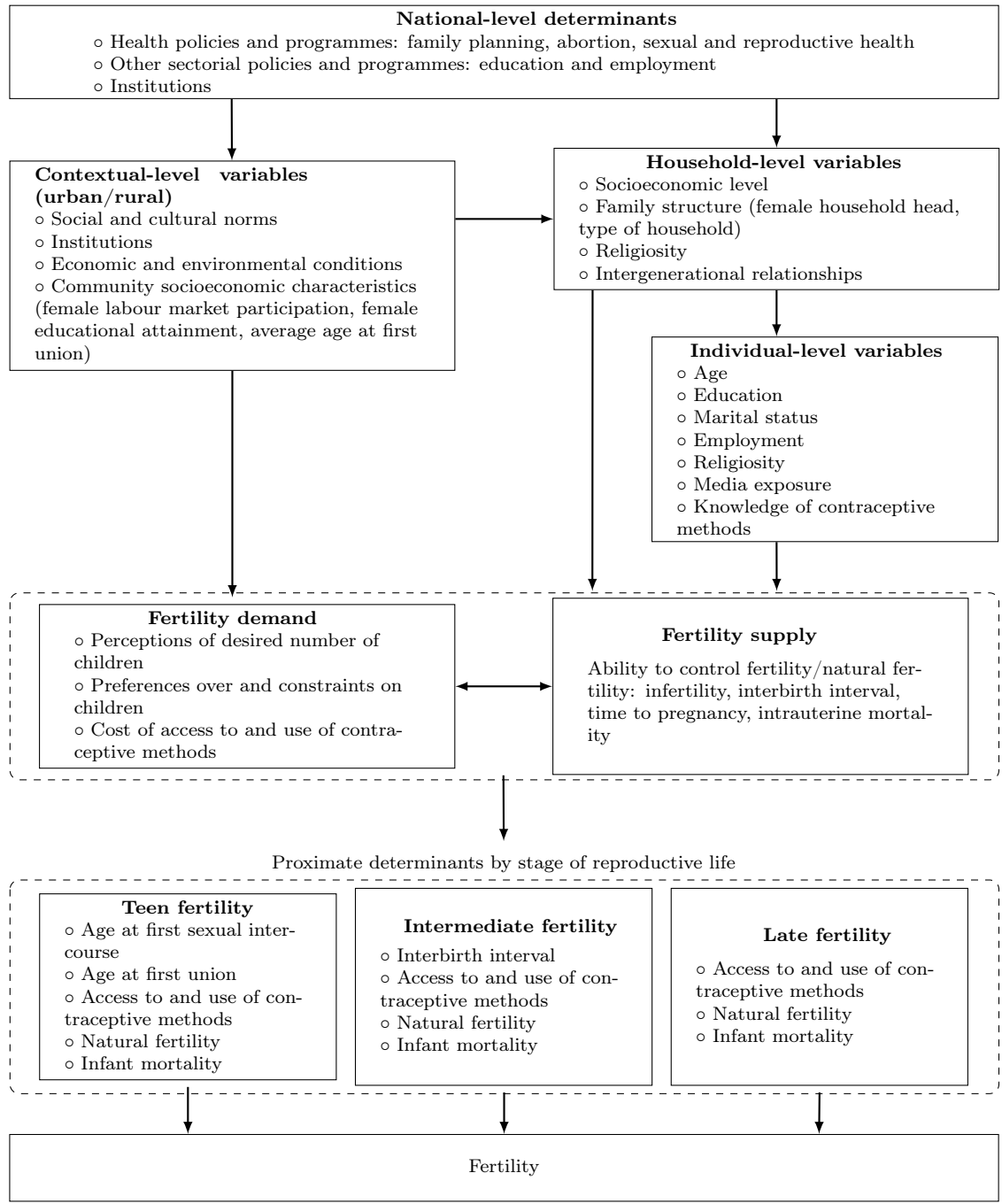

First, the demographic transition theory \parencite{farooq1985, un1973} describes the shift from a population with high fertility and mortality to one with low fertility and mortality as a result of economic development. Originally used to explain the demographic change in Great Britain during the Industrial Revolution, it postulates that fertility decline is a consequence of a country's modernisation, economic growth and development, with the reduction in infant mortality and rural-to-urban migration being the main drivers. 

Next, the so-called conventional structural hypothesis derives not only from the demographic transition theory but also microeconomic models and the threshold hypothesis. The former are the result of applying economic analysis, with rational choice as the main workhorse, to fertility, particularly to explain families' preferences for having children \parencite{becker1973,easterlin1969,leibenstein1975}. This body of literature develops a supply--demand theoretical framework for fertility, where this variable is the result of the supply of children (number of surviving children in the absence of birth control), demand for children (due to preferences about the number of offspring) and cost of regulating births. The latter, the threshold hypothesis \parencite{un1963}, states that fertility will decline only after socioeconomic and health conditions have reached a certain level.

Third, the ideational or diffusionist hypothesis posits that the evolution of fertility is due to shifts in perceptions of, ideas about and attitudes towards birth control. These changes have their roots in the expansion or diffusion of family planning organizations and mechanisms and the increase in women's educational attainment. However, many authors emphasise that the operation of these forces requires the prior achievement of a certain level of socioeconomic development \parencite{caldwell1992, cleland1987, cleland2001, fort2016, hirschman1990}. 

Fourth, the relationship between fertility and female labour market participation has also received much attention from researchers. In particular, the so-called maternal role incompatibility hypothesis focuses on disentangling the potential and eventual problems of reconciling preferences over the number of children with working life prospects \parencite{cramer1980, lehrer1986, spitze1988}. Similarly, the societal response hypothesis argues that the existence of policies aimed at minimising conflicts between motherhood and female labour market participation (e.g., available and affordable childcare, generous parental leave or changes in attitudes towards working mothers) could prevent an increase in female employment from translating into a decline in fertility \parencite{brewster2000, rindfuss2003}.

The final approach is the institutional perspective, which focuses on the context, in particular the institutions, shaping fertility decisions. This perspective aims to reconcile the macro- and microeconomic perspectives.

Bridging the theoretical insights summarised above and empirical practice is challenging. It requires a search for variables that adequately approximate the different dimensions suggested by the theory. To capture the socioeconomic dimension, the variables most commonly chosen in the literature are GDP per capita or household expenditure \parencite{buckles2021,chatterjee2016,sobotka2011}; unemployment \parencite{cazzola2016,currie2014}; development indicators such as the Gini index, basic infrastructure and services and health and education expenditure \parencite{bettio1998,engelhardt2004b,engelhardt2004a}; female educational attainment (either enrolment rates or average years of schooling by cohort) \parencite{ainsworth1996,sackey2005,schultz1973,vavrus2003} and labour market indicators (women's labour market participation, female wages and the gender pay gap) \parencite{kato2021}. 

The main proxies for the cultural dimension are age of sexual initiation, availability and use of contraceptive methods and household time allocation. The most widely used variables to approximate demographic aspects are population structure and mortality (total or infant). Finally, to operationalize institutional aspects and public policies, which cut across all the other dimensions, a popular strategy is to specify main milestones in the implementation of or drastic changes in public interventions, among others, related to family planning or education \parencite{carr2017,kearney2015a,paton2020}. 

The previous empirical literature to which this study refers includes both time series and panel data analyses. The former type of research tends to highlight the role of macro-level determinants of fertility. Specifically, these studies emphasise the relevance of female labour force participation, unemployment (both male and female), infant mortality and women's education, among other factors. Nevertheless, this literature is inconclusive with respect to these determinants. A consensus is lacking regarding whether the relationships are causal (even in Granger's sense) or whether they could even be bidirectional \parencite{chatterjee2016,sobotka2011,kato2021,audi2021}.

The evidence on the effect of GDP per capita is more complex. Fertility appears to be procyclical, but it also tends to fall in the long term with economic growth and in the short run with recessions \parencite{audi2021,chatterjee2016,sobotka2011}. The effect also differs across reproductive life stages, with the fertility of women aged 30 and over being the most sensitive to economic fluctuations. The related literature even discusses whether fertility actually declines after economic crises or, in contrast, whether such a development actually precedes the recorded output losses since it is extremely dependent on short-term expectations, as suggested by \textcite{buckles2021} for the United States. For this reason, i.e., the anticipatory behaviour of fertility, these authors are quite critical of the use of unemployment and other business cycle indicators as explanatory factors for fertility. By contrast, other studies such as \textcite{currie2014}, which link fertility and unemployment by cohort, suggest an important role for labour market prospects. In particular, they find that women aged 20--24 are the most affected group and that the negative impact increases over time due to the effect on childless women. 

Previous works also make use of other measures of economic performance, such as women's wage levels, female labour market participation and the gender pay gap. For instance, \textcite{kato2021}, using data from 1980 to 2019 for Japan, with its low fertility rate and labour shortage (circumstances very far from the Uruguayan reality), suggests that women's average earnings have a negative impact on childbirth. Their results highlight the importance of the opportunity cost of having children and the need to design policies that improve work--life balance.

The inclusion of female education makes it possible to test the validity of the diffusion hypothesis. By way of example, the work of \textcite{audi2021}, employing time series from 1971 to 2014 for Tunisia, finds a negative impact of women's schooling level on fertility.

The United States followed a very similar path to Uruguay’s. Although the decline was not monotonic, its total fertility rate more than halved between 1900 and 2017, and teenage fertility did not decline until recently. \textcite{buckles2019} examine the heterogeneity in fertility trends across different demographic groups. They find that the reduction in fertility in recent decades followed the reproductive behaviour of young and single women whereas married women and those above 30 saw an increase in their fertility. These authors’ results also confirm the positive correlation between declines in fertility and in the proportion of unplanned pregnancies. 

The literature using panel data exploits differences either between regions within the same national boundaries or among countries to shed light on the main determinants of fertility. Regarding the former, we can highlight the study by \textcite{kearney2015b} for the United States, whose main finding is the role played by the expansion in access to family planning services, which explains 13\% of the drop in teenage fertility between 1991 and 2010. The work of \textcite{cazzola2016} for Italy emphasises the importance of unemployment, especially in the case of male rates. The Japanese case has also received some attention. \textcite{suzuki2019} finds that female wages have a nonnegligible negative impact on fertility whereas, according to the analysis of \textcite{kato2021}, differences in birth rates are due to childcare availability and female labour force participation.

Regarding cross-country literature, most of the existing works centre on developed countries. For instance, \textcite{sobotka2011} confirm the negative impact of economic crises on fertility rates. In terms of policies, \textcite{addio2006} show that social transfers that reduce the direct cost of children and provisions that allow mothers to better balance work and family have a significant impact on birth rates. The analysis of \textcite{kato2021} comes to similar conclusions with regard to the economic environment but, surprisingly, opposite ones for family policies. The author also stresses the importance of female labour market conditions. Interestingly, \textcite{paton2020} find no effect of the expansion of sexual education on fertility. Among the studies focusing on developing regions, we can mention the work of \textcite{ojakaa2022} for sub-Saharan Africa, which emphasizes the role of age at first marriage and contraceptive prevalence. For Latin America, \textcite{palloni1999} analyse the relationship between infant mortality and fertility rates between 1920 and 1990 and find small positive effects of infant mortality on fertility.

The contribution of this work to the existing literature is twofold. First, it is the first study to focus specifically on Uruguay. This country has idiosyncratic characteristics: it shares many features with Latin America and the Caribbean (such as relatively high levels of inequality and labour market informality), but its level of social development is historically high, it reached the high-income country category less than a decade ago, and it was one of the first states in the hemisphere to complete its demographic transition (while teenage fertility remained high until very recently). We are not aware of any other research work exclusively devoted to a Latin American or Caribbean country. Second, by using the most recent and comprehensive data at the national and departmental level and state-of-the-art econometric techniques, we expand the empirical evidence on the main drivers of fertility at different stages of reproductive life. Thus, our study also aims to contribute to a better understanding of the overall dynamics of birth rates. 

\section{Time series country-level analysis}\label{Section 3}
\subsection{Data and methods}\label{Section 3.1}

In this section, we carry out a descriptive analysis of the evolution of fertility at different stages of reproductive life (adolescents, women aged 20 to 29 and women aged 30 and over) based on historical time series. To this end, we collect information on the following covariates: GDP per capita, female employment rate (share of employed women in relation to female working-age population), female secondary gross enrolment rate (number of women in secondary education as a percentage of women aged 12--17) and infant mortality rate (number of deaths of children under one year of age, expressed per 1,000 live births). We use GDP per capita as a proxy for socioeconomic development. Our consideration of the female employment rate and secondary school enrolment allows us to test the validity of the diffusion hypothesis. These variables capture shifts in women's preferences over and attitudes towards fertility. In particular, educational attainment might improve women’s access to information on contraceptive methods and increase their intrahousehold bargaining power. Infant mortality rate plays a key role in the demographic transition hypothesis and in other modern population theories. The relationship between fertility and infant mortality can be difficult to unravel because of various underlying mechanisms. These channels can range from purely physiological effects, where the death of an infant triggers resumption of mothers’ menstruation and ovulation, thus increasing their likelihood of a new conception, to replacement or insurance mechanisms (whereby families aim for a specific number of surviving children beyond the desired family size), distortions in the market for potential partners and competition between children for maternal care and household resources \parencite{palloni1999, wolpin1998}. Additionally, infant mortality can serve as an indicator of the quality of a country's health systems and level of access to medical care.

The availability of information on fertility and covariates over time varies considerably. Overall, in our econometric exercise, we are able to analyse teenage fertility over the period 1968--2021 and the rates of the other two age groups from 1978 to 2021. Reconstructing the time series of these variables is not trivial, requiring substantial effort and the combination of different sources. First, the adolescent fertility rate comes from statistical information from the Uruguayan National Institute of Statistics \parencite{ine2023a} and the World Development Indicators (WDI) \parencite{wb2023}, whereas we obtain the other two fertility rates by combining vital statistics from the Ministry of Public Health and population projections from the National Statistics Institute \parencite{ine2023b,msp2023}. We retrieve historical data on GDP per capita (in constant 1990 US\$) from the Montevideo--Oxford Latin American Economic History Data Base \parencite{moxlad2023}, and the female high school enrolment rate comes from the WDI \parencite{wb2023}. We reconstruct our series on the female employment rate using information from \textcite{claeh1990} and \textcite{ine2023c}.\footnote{The Uruguayan national household survey is not nationally representative until 1995. Nevertheless, since its inception, it has covered all municipalities with 5,000 inhabitants. Therefore, to construct homogeneous series, we restrict all the statistical information based on this database to the mentioned localities.} Finally, we obtain the infant mortality rate for the period of interest from \textcite{ine2023d}. Table~\ref{Table 1} shows the descriptive statistics of the variables included in our econometric exercise. 

\begin{singlespace}
\begin{table}[!ht]
\begin{ThreePartTable}
\def\sym#1{\ifmmode^{#1}\else\(^{#1}\)\fi}
\footnotesize
\setlength\tabcolsep{0.2em}
\begin{TableNotes}[flushleft]\setlength\labelsep{0pt}\footnotesize\justifying
\item\textit{Note}: The number of observations is 54 (1968--2021) in all cases except for the 20--29 and 30+ fertility rates, where it is 44 (1978--2021). 
\end{TableNotes} 
\begin{tabularx}{\textwidth}{X *{4}{S[table-column-width=2cm]}}
\caption{Summary statistics of time series data} \label{Table 1}\\
\toprule
&\multicolumn{1}{c}{Mean}&\multicolumn{1}{c}{\makecell{Standard\\deviation}}&\multicolumn{1}{c}{Minimum}&\multicolumn{1}{c}{Maximum}\\
\midrule
Fertility rate 15--19&61.9&10.8&25.9&74.0\\
Fertility rate 20--29&110.7&24.8&61.4&150.8\\
Fertility rate 30--49&41.6&4.5&31.7&48.7\\
GDP per capita&7982.4&2711.2&4747.2&13267.1\\
Female employment rate&89.7&22.0&61.7&131.3\\
Female high school enrolment&38.8&9.1&23.1&52.4\\
Infant mortality rate&23.5&15.7&6.2&61.9\\
\bottomrule
\insertTableNotes 
\end{tabularx}
\end{ThreePartTable}
\end{table}
\end{singlespace}
\FloatBarrier

\begin{figure}[!ht]
\footnotesize
\caption{Evolution of age-specific fertility rates in Uruguay (births per 1,000 women, 1976--2021)}
\centering 
\includegraphics[width=0.8\textwidth]{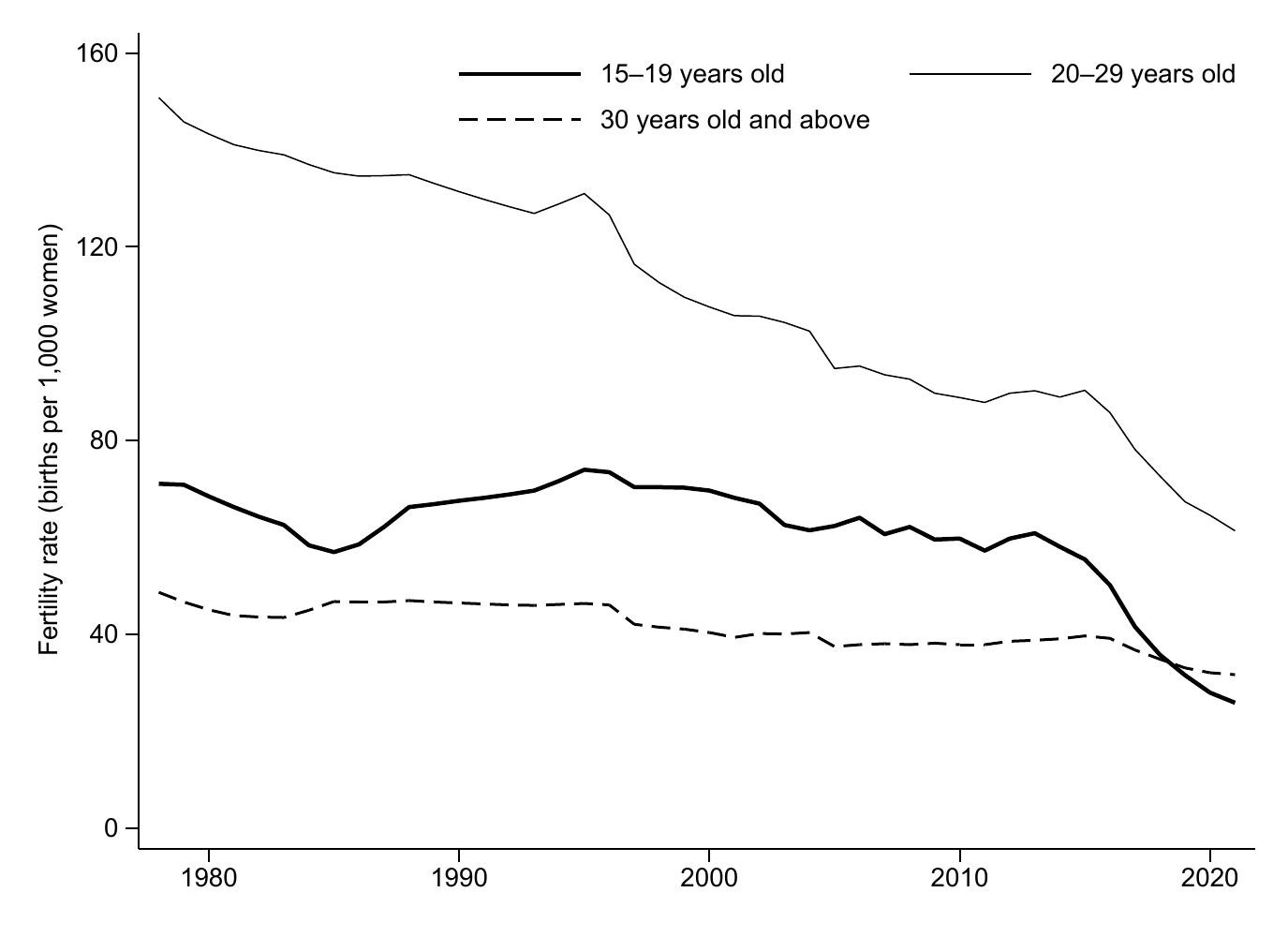}\\
\justifying
\noindent\textit{Source:} Authors' analysis from \textcite{ine2023a} and \textcite{wb2023}. 
\label{Figure 2}
\end{figure}

Figure~\ref{Figure 2} shows the evolution of the fertility rate by age group from 1978 to 2021 (the time window for which we have information on all of the groups). Different patterns emerge. First, the adolescent birth rate remained relatively high until 2015, when it experienced a rather abrupt fall. Second, the fertility rate of women between 20 and 29 years old underwent a sustained declined throughout the whole analysed period that accelerated in 2016. Last, the fertility of women aged 30 and above decreased at a much slower pace over the more than four decades considered in the analysis. 

Naturally, the period covered by our analysis witnessed the enactment of several potentially relevant education- and health-related laws. Unfortunately, the degrees of freedom of our research design and the fact that some of the developments were contemporaneous prevent us from disentangling the causal effect of these policies. Below, we discuss the inclusion of a linear time trend and dummies for certain subperiods of time.

\FloatBarrier
In principle, we aim to estimate the effect of the covariates of interest through the following linear model:
\begin{equation}
\label{eq 1}
y_t = \mu + \delta' X_{t} + \xi_t
\end{equation}
where $y_t$ is the fertility rate (in natural logs), $X_t$ represents a vector comprising the covariates mentioned above (in natural logs), and $\epsilon$ denotes the random disturbance term.

To proceed with the time series analysis, first of all, we must check whether the variables included in the analysis are stationary using the augmented Dickey--Fuller (ADF) \parencite{dickey1981} and Phillips--Perron (PP) \parencite{phillips1988} unit-root tests. The latter is robust to heteroscedasticity and serial correlation and does not require specification of the form of the lag structure. 

Second, having established the stationarity of the series, we examine the existence of cointegration between fertility and the variables described above using the tests proposed by \textcite{engle1987}, \textcite{johansen1995b}, \textcite{pesaran1999} and \textcite{pesaran2001}.

Previous tests have been shown to be highly sensitive to the chosen model specifications. In third place, to address this issue, we decide to employ the autoregressive distributed lag (ARDL) model in our investigation of cointegration \parencite{pesaran1999, pesaran2001}:
\begin{equation}
\label{eq 2}
y_t = \alpha_0 + \alpha_1 t + \sum_{i=1}^{p} \phi_i y_{t-1} + \sum_{i=0}^{q} \beta'_{i} X_{t-i} + \epsilon_t
\end{equation}
where $t$ is a linear time trend, $\epsilon$ denotes the random disturbance term and $p$ and $q$ are the number of lags of the dependent and independent variables, respectively. In practice, we allow for a different structure of each variable (so that $q$ can vary for each of the four covariates). We use the Bayesian information criterion (BIC) to determine the optimal number of lags (which may be different for each variable). The advantage of this model over other approaches is that it allows for mixed orders of cointegration and performs better with small samples.

Fourth, in the case of evidence of cointegration, we can rewrite equation~\ref{eq 2} as
\begin{equation}
\begin{aligned}
\label{eq 3}
\Delta y_t ={} & \alpha_0 + \alpha_1 t - \gamma \left(y_{t-1} - \theta' X_{t-1}\right) + \sum_{i=1}^{p-1} \psi_{y_i} \Delta y_{t-1} + \omega'\Delta X_t \\
&+ \sum_{i=1}^{q-1} \psi'_{X_i} \Delta X_{t-i} + u_t
\end{aligned}
\end{equation}
where $\theta$ denotes the long-run coefficients (the equilibrium effects of the covariates on fertility); $\gamma$ represents the error correction term (ECT)---the (negative) speed-of-adjustment coefficient, which measures how fast the dependent variable responds to deviations from the equilibrium relationship; and $\psi_{y_i}$, $\omega$ and $\psi_{X_i}$ capture short-term fluctuations (unrelated to the long-term equilibrium).

Finally, we carry out goodness-of-fit tests on the ARDL model, including tests for first-order autocorrelation \parentext{Breusch--Godfrey \parencite{breusch1978, godfrey1978} and Durbin--Watson \parencite{durbin1950,durbin1951,durbin1971}}, heteroscedasticity \parentext{\textcite{white1980}, Breusch--Pagan \parencite{breusch1979} and Cook--Weisberg \parencite{cook1983}}, and normality \parencite{dagostino1990}. In addition, we examine \textcite{granger1969} causality and compute impulse response functions and the error variance decomposition.

\subsection{Results}\label{Section 3.2}

According to the unit-root tests described above, whose results we present in the appendix (Table~\ref{Table A1}), the variables included in our model are first-difference stationary---$I(1)$. We allow for a maximum of two lags because of the limited statistical power due to our sample size and number of variables. The results of the tests for cointegration (Tables~\ref{Table A2}--\ref{Table A4}) indicate the existence of at least one cointegration relationship.

On the other hand, as mentioned earlier, the ARDL model for each series indicates the optimal lag structure to be as follows: (2,0,0,0,2) for teen fertility, (2,1,0,0,1) for intermediate fertility and (1,2,0,0,1) for late fertility.

In Table~\ref{Table A5}, we present the results of the goodness-of-fit tests. They indicate that we cannot reject the null hypotheses of absence of serial correlation, homoscedasticity, and normality.

To check the stability of the parameters in our econometric models, we perform cumulative sum of squares tests for structural change \parencite{brown1975}. The plot shown in Figure~\ref{Figure A1} indicates that the cumulative sum of the squared recursive residuals is approximately within the 95\% confidence interval for the target value based on the null hypothesis of the parameter at each point all the time for the three fertility rates.

Since we observe relevant changes in the late fertility rate in 1996 and 2004 and an abrupt fall in all three rates from 2016, we include three dummy variables to account for these changes and ensure the stationarity of the time series.

Table~\ref{Table 2} presents the main results of our analysis. They show the existence of a long-term relationship between fertility and the covariates, with a statistically significant negative speed of adjustment. The mentioned error correction term indicates that the fertility rate adjusts to temporary deviations at a rate of between 17.5\% and 25.4\% per year, depending on the age group. Regarding the estimated long-run coefficients, which we interpret as elasticities, first, GDP per capita has a statistically significant effect, with a negative impact, on fertility only among women aged 30 and over. Second, the impact of the female employment rate is statistically different from zero and positive in all cases. A 1\% increase in the share of employed women raises fertility by between 0.229\% (women aged 20--29) and 0.475\% (women aged 30 and over). Third, female high school enrolment exerts a statistically significant and positive effect on adolescent fertility and a negative one on that of women aged 30 and over. Fourth, infant mortality rate is relevant only for the latter group of women, with a positive statistically significant impact. 

The existence of discrepancies between the short- and long-run coefficients simply indicates the complexity of the dynamic interactions between fertility and the covariates. Regarding adolescent fertility, we can interpret the short-term relationships as indicative of the necessary initial conditions for curbing this age-specific rate. Namely, a decline in teenage fertility requires an increase in women's educational attainment. Specifically, a 1\% rise in female secondary school enrolment reduces the adolescent fertility rate by 0.379\% with a two-year lag. However, this variable has a positive statistically significant effect for the other two age groups (0.143 for women aged 20--29 and 0.146 for women aged 30 and over). Nevertheless, this finding is consistent with the evidence reported by \textcite{fort2016} for England and continental Europe. For the former, these authors find support for a negative relationship between education and total fertility. This effect does not hold for mainland Europe. These authors suggest that this discrepancy might be due to the higher adolescent birth rate in England, where the increase in educational attainment associated with the expansion of compulsory schooling exerted an almost mechanical negative effect on fertility. 

Finally, as mentioned above, the relationship between GDP per capita and fertility is far from simple, and in the short term, it may be the opposite of the inverse relationship that one expects in the long run. For example, economic or social crises may temporarily increase fertility due to uncertainty and the need for family support or in response to policies that encourage childbearing. In the long term, however, one anticipates an inverse relationship, associated with better access to education, increased employment opportunities and improved health care services. The effect also differs according to the stage of reproductive life, with the late stage being the most sensitive to economic fluctuations, as observed in Table~\ref{Table 2}. For women aged 30 and over, we find a significant and negative relationship at time $t$ ($-0.171$) and a positive one at $t-1$ ($0.153$).

The coefficient of the temporal dummy variable 2016--2021 accounts for a decrease of 0.110\% in the adolescent fertility rate and of 0.049\% in the intermediate fertility rate during the period of interest. We believe that this variable may capture changes that occurred in those years or in previous years. Uruguay launched several public policies that could potentially affect fertility, especially adolescent fertility, such as the following: the Sexual and Reproductive Health Law (2008) \parencite{ley18246}, an expansion of the range of available contraceptives (including subdermal implants) from 2015 onward, and the creation of a network of sexual health service providers and the expansion of reproductive health services, including spaces for adolescents and the creation of a strategy for the prevention of unwanted adolescent pregnancies. The dummy variable is not significant for the late fertility rate, which may simply reflect that the aforementioned policies mainly targeted other groups. Regarding other temporal variables, a negative and significant trend in the adolescent fertility rate ($-0.010$) and an increase in subsequent fertility rates in 2004 ($0.064$) stand out. The latter could have to do with the economic growth after the 2002 crisis. For instance, Uruguay's GDP grew by 11.1\% in 2004 \parencite{ine2023g}.

\begin{singlespace}
\begin{table}[!ht]
\begin{ThreePartTable}
\def\sym#1{\ifmmode^{#1}\else\(^{#1}\)\fi}
\footnotesize
\begin{TableNotes}[flushleft]\setlength\labelsep{0pt}\footnotesize\justifying
\item\textit{Notes}: Standard errors in parentheses. All the models include an intercept. The structures of the three models in the table are ARDL(2,0,0,0,2), ARDL(2,1,0,0,1) and ARDL(1,2,0,0,1), respectively. \sym{***} significant at 1\%; \sym{**} significant at 5\%; \sym{*} significant at 10\%.
\end{TableNotes}
\begin{tabularx}{\textwidth}{X *{3}{S[table-column-width=2cm]}}
\caption{Estimation results of the ARDL model} \label{Table 2}\\
\toprule
&\multicolumn{1}{c}{(I)}&\multicolumn{1}{c}{(II)}&\multicolumn{1}{c}{(III)}\\[1ex]
&\multicolumn{3}{c}{Fertility rate (in logs) of women aged}\\[1ex]
&\multicolumn{1}{c}{15--19}&\multicolumn{1}{c}{20--29}&\multicolumn{1}{c}{30 and above}\\
\midrule
Error correction term&-0.175\sym{***}&-0.254\sym{**}&-0.194\sym{**}\\
&(0.039)&(0.083)&(0.942)\\
Long-run relationships &&&\\[0.5ex]
~~$\log \left(\text{GDP per capita}\right)_t$&0.060&-0.022&-0.165\sym{***}\\
&(0.045)&(0.051)&(0.047)\\ 
~~$\log \left(\text{Female employment rate}\right)_t$&0.302\sym{***}&0.229\sym{*}&0.475\sym{***}\\
&(0.076)&(0.099)&(0.108)\\ 
~~$\log \left(\text{Female high school enrolment}\right)_t$&0.295\sym{***}&0.140&-0.171\sym{***}\\
&(0.068)&(0.068)&(0.055)\\ 
~~$\log \left(\text{Infant mortality rate}\right)_t$&0.054&0.054&0.080\sym{*}\\
&(0.050)&(0.050)&(0.047)\\ 
Short-run relationships &&&\\[0.5ex]
~~$\Delta \left(\log \left(\text{Age-specific fertility rate}\right)\right)_{t-1}$ &0.454\sym{***}&0.320\sym{*}&\\
&(0.112)&(0.165)&\\ 
~~$\Delta \left(\log \left(\text{GDP per capita}\right)\right)_{t}$&&-0.127\sym{*}&-0.171\sym{**}\\
&&(0.069)&(0.073)\\
~~$\Delta \left(\log\left(\text{GDP per capita}\right)\right)_{t-1}$&&&0.153\sym{*}\\
&&&(0.079)\\
~~$\Delta \left(\log\left(\text{Female high school enrolment}\right)\right)_{t}$&0.050&0.143\sym{*}&0.146\sym{*}\\
&(0.091)&(0.078)&(0.082)\\
~~$\Delta \left(\log\left(\text{Female high school enrolment}\right)\right)_{t-1}$ &-0.379\sym{***}&&\\ 
&(0.091)&&\\
Year 1996 &&&0.021\\
&&&(0.019)\\
Year 2004 &&& 0.064\sym{***}\\
&&&(0.021)\\ 
Years 2016--2021&-0.110\sym{***}&-0.049\sym{***}&0.007\\
&(0.025)&(0.014)&(0.014)\\
Linear time trend&-0.009\sym{***}&-0.004&0.002\\
&(0.003)&(0.004)&(0.003)\\ [1ex] 
No. of observations&52&42&43\\
Mean of dependent variable&61.93&119.68&41.56\\
ARDL model structure&\multicolumn{1}{c}{(2,0,0,0,2)}&\multicolumn{1}{c}{(2,1,0,0,1)}&\multicolumn{1}{c}{(1,2,0,0,1)}\\ 
\bottomrule 
\insertTableNotes
\end{tabularx}
\end{ThreePartTable}
\end{table}
\end{singlespace}

According to \posscite{granger1969} causality criterion, we detect a bidirectional relationship between the adolescent fertility rate and the secondary school enrolment rate. Table~\ref{Table 3} shows that, in the past, the education variable was able to predict, in the Granger sense, the current rate of adolescent fertility and vice versa. This underlines the relevance of education as a policy measure to influence the adolescent fertility rate but not fertility across reproductive life stages. 

Finally, to understand the relative importance of each factor in explaining the variability of the time series, we estimate impulse response functions (IRFs) of both the fertility rate and its determinants. IRFs describe the dynamic response of a system to a shock or impulse. It shows how the fertility reacts over time to a sudden change in one of its inputs. This information can be used to understand the transmission of shocks or policy interventions. Figure~\ref{Figure A2} in the appendix shows that an increase in high school enrolment leads to a decrease in fertility in the short run but that the effect is relatively small and diminishes over time. However, the largest response corresponds to changes in fertility itself. These can be associated, for example, with shocks to fertility preferences or to fertility behaviour.\footnote{In principle, in orthogonal IRFs, the results depend on the order in which one includes the variables in the model. In practice, however, they are similar in all cases in our analyses, irrespective of the order.} 

As an alternative measure to the IRFs, we present the forecast error variance decomposition with different time horizons for each model in Tables~\ref{Table A6}--\ref{Table A8}. The IRFs provide information on the dynamic response of the system but do not reveal the sources of variability in the time series. The main salient finding is that more than 75\% of the variation in fertility in the long run is due to its own shocks rather than to its determinants. Anyway, it is worth highlighting the contribution of education by age group. Secondary school enrolment accounts for 10.8\% of the variation in adolescent fertility in the first period, and its effect vanishes over time. In contrast, for women aged 20--29 and those aged 30 and over, the initial contribution is very low but grows over time (more than 8\% after eight time periods). 

\begin{landscape}
\begin{singlespace}
\begin{table}[!ht]
\begin{ThreePartTable}
\def\sym#1{\ifmmode^{#1}\else\(^{#1}\)\fi}
\footnotesize
\setlength\tabcolsep{0.2em}
\begin{TableNotes}[flushleft]\setlength\labelsep{0pt}\footnotesize\justifying
\item\textit{Notes}: The degrees of freedom refer to the number of constraints in the model (two in the case of the rows due to individual variables and eight in the case of the ones due to all variables).
\end{TableNotes}
\begin{tabularx}{\linewidth}{p{0.28\linewidth}X *{6}{S[table-column-width=1.35cm]}}
\caption{Results of the Granger causality test} \label{Table 3}\\
\toprule
\multicolumn{2}{l}{\multirow{3}{*}{Equation}}&\multicolumn{6}{c}{Women aged}\\[0.5ex]
&&\multicolumn{2}{c}{15--19}&\multicolumn{2}{c}{20--29}&\multicolumn{2}{c}{30 and above}\\[0.5ex]
&&\multicolumn{1}{c}{$\chi^2$}&\multicolumn{1}{c}{\textit{p}-value}&\multicolumn{1}{c}{$\chi^2$}&\multicolumn{1}{c}{\textit{p}-value}&\multicolumn{1}{c}{$\chi^2$}&\multicolumn{1}{c}{\textit{p}-value}\\ 
\midrule
$\Delta \log\left(\text{Age-specific fertility rate}\right)$&$\Delta \log\left(\text{GDP per capita}\right)$&15.615&0.458&33.945&0.183&42.661&0.118\\
&$\Delta \log\left(\text{Female employment rate}\right)$&17.935&0.408&21.669&0.338&38.644&0.145\\
&$\Delta \log\left(\text{Female high school enrolment}\right)$&19.616&0.000&40.911&0.129&37.947&0.150\\ 
&$\Delta\log\left(\text{Infant mortality rate}\right)$&21.393&0.343&0.431&0.806&0.525&0.769\\ 
&All variables&36.464&0.000&14.113&0.079&11.990&0.152\\[1ex] 
$\Delta \log\left(\text{GDP per capita}\right)$&$\Delta \log\left(\text{Age-specific fertility rate}\right)$&0.547&0.761&0.464&0.793&2.256&0.324\\
&$\Delta \log\left(\text{Female employment rate}\right)$&0.172&0.918&0.126&0.939&0.406&0.816\\
&$\Delta \log\left(\text{Female high school enrolment}\right)$&0.686&0.710&0.652&0.722&0.699&0.705\\ 
&$\Delta \log\left(\text{Infant mortality rate}\right)$&16.503&0.438&14.417&0.486&14.225&0.491\\ 
&All variables&37.333&0.880&32.151&0.920&5.123&0.744\\[1ex] 
$\Delta \log\left(\text{Female employment rate}\right)$&$\Delta \log\left(\text{Age-specific fertility rate}\right)$&25.952&0.273&21.591&0.340&13.927&0.498\\
&$\Delta \log\left(\text{GDP per capita}\right)$&12.059&0.002&46.594&0.097&45.657&0.102\\
&$\Delta \log\left(\text{Female high school enrolment}\right)$&48.206&0.090&55.275&0.063&52.908&0.071\\ 
&$\Delta \log\left(\text{Infant mortality rate}\right)$&43.433&0.114&36.131&0.164&45.316&0.104\\ 
&All variables&19.084&0.014&12.498&0.130&11.552&0.172\\[1ex]
$\Delta \log\left(\text{Female high school enrolment}\right)$&$\Delta \log\left(\text{Age-specific fertility rate}\right)$&74.073&0.025&21.333&0.344&22.475&0.325\\
&$\Delta \log\left(\text{GDP per capita}\right)$&16.055&0.448&0.213&0.899&0.185&0.912\\
&$\Delta \log\left(\text{Female employment rate}\right)$&18.124&0.404&2.603&0.272&23.554&0.308\\ 
&$\Delta \log\left(\text{Infant mortality rate}\right)$&0.765&0.682&26.545&0.265&24.514&0.294\\ 
&All variables&16.228&0.039&93.533&0.313&94.861&0.303\\[1ex]
$\Delta \log\left(\text{Infant mortality rate}\right)$&$\Delta \log\left(\text{Age-specific fertility rate}\right)$&18.037&0.406&72.559&0.027&31.317&0.209\\
&$\Delta \log\left(\text{GDP per capita}\right)$&35.062&0.173&16.784&0.432&0.592&0.744\\
&$\Delta \log\left(\text{Female employment rate}\right)$&0.494&0.781&0.969&0.616&1.575&0.455\\
&$\Delta \log\left(\text{Female high school enrolment}\right)$&0.547&0.761&1.200&0,549&0.337&0.845\\ 
&All variables&92.138&0.325&16.008&0.042&11.151&0.193\\ 
\bottomrule 
\insertTableNotes
\end{tabularx}
\end{ThreePartTable}
\end{table}
\end{singlespace}
\end{landscape}

\FloatBarrier
\section{Panel data department-level analysis}\label{Section 4}
\subsection{Data and methods}\label{Section 4.1}

In this section, we make use of panel data for the country's 19 administrative units (departments) and the period 1984--2019 to provide additional insight into the dynamics of fertility in Uruguay. This strategy allows us to increase the statistical power of the analysis and control for time-constant department-level heterogeneity through fixed effects techniques, thereby mitigating endogeneity problems.

It is worth noting the existence of significant territorial disparities within the country. The department of Montevideo (which includes the country capital of the same name) concentrates 40\% and 50\% of the national population and GDP, respectively \parencite{otu2023}. From a multidimensional perspective, although the human development index (HDI) experienced sustained progress across the whole country from 2008 to 2018, again, a nonnegligible gap between Montevideo and the rest of Uruguay is observable \parencite{otu2023}. In 1998, Montevideo was the only department to exhibit very high human development (above 0.800), while the rest of the regions had high HDI values (between 0.700 and 0.800). In 2018, apart from Montevideo, four other departments (Colonia, Maldonado, Flores and Florida) had crossed the threshold of very high human development. Cerro Largo, Rivera, Rocha, Treinta y Tres, Tacuaremb{\'o} and Artigas were, in descending order, the areas with the lowest HDI values in Uruguay.

To build our database, we rely on a variety of sources, balancing the convenience of long series with the availability of statistical information. Whereas we can include a larger number of variables in this analysis than in our time series econometric exercise, the period covered here is shorter. As in the previous section, we model age-specific fertility rates as a linear function of several covariates. We compute department-level fertility rates from vital statistics \parencite{msp2023} and population projections \parencite{ine2023f}. First, the covariates include the demographic structure of the department through the percentage of each age group of total women aged 15--49, calculated from population projections \parencite{ine2023f}. The second variable, derived from the national household survey \parencite{ine2023e}, is the percentage of women in each age group married or in a union. To assess the impact of education, we consider a demand-side indicator, the average years of schooling of women in each age group, derived from \parencite{ine2023e}. Fourth, using the same data source, we consider a set of variables that aim to capture the opportunity cost of having children, such as age-specific female employment rates. In fifth place, we also consider the gender gap, calculated as the ratio of women's average labour income to men's average earnings. Furthermore, our analysis considers average household disposable income per capita in national currency units (NCUs) at December 2010 prices. The infant mortality rate is computed from \textcite{msp2023}. Table~\ref{Table 4} shows the summary statistics of the variables included in our analyses.\footnote{As in the time series analysis, to rely on homogenous series, we limit our analysis of the Uruguayan household survey to municipalities with 5,000 inhabitants or more.}

\begin{landscape}
\vspace*{\fill}
\begin{singlespace}
\begin{table}[!ht]
\begin{ThreePartTable}
\def\sym#1{\ifmmode^{#1}\else\(^{#1}\)\fi}
\footnotesize
\setlength\tabcolsep{0.3em}
\begin{TableNotes}[flushleft]\setlength\labelsep{0pt}\footnotesize\justifying 
\item\textit{Note}: The number of observations is 665 (19 departments from 1984 to 2019).
\end{TableNotes} 
\begin{tabularx}{\linewidth}{X *{6}{S[table-column-width=2cm]}}
\caption{Summary statistics of regional panel data} \label{Table 4}\\
\toprule
&\multicolumn{1}{c}{Mean}&\multicolumn{1}{c}{\makecell{Standard\\deviation}}&\multicolumn{1}{c}{Minimum}&\multicolumn{1}{c}{Maximum}&\multicolumn{1}{c}{\makecell{Between-region\\standard\\deviation}}&\multicolumn{1}{c}{\makecell{Within-region\\standard\\deviation}}\\
\midrule
Fertility rate 15--19&70.8&17.4&18.5&112.4&11.2&13.6\\
Fertility rate 20--29&123.0&26.7&55.2&186.9&11.5&24.3\\
Fertility rate 30+&42.0&8.1&24.8&71.0&3.8&7.2\\
\% of women 15--19&14.3&0.9&12.0&16.5&0.7&0.6\\
\% of women 20--29&25.0&1.4&21.7&29.3&1.0&1.0\\
\% of women 30--49&45.3&2.3&37.9&50.2&1.5&1.7\\
\% of women married 15--19&1.9&0.9&0.0&5.6&0.4&0.8\\
\% of women married 20--29&22.1&3.2&10.2&32.0&1.4&2.9\\
\% of women married 30--49&76.0&3.5&65.8&89.1&1.7&3.1\\
Average years of schooling of women 15--19&8.9&0.4&7.8&10.2&0.4&0.8\\
Average years of schooling of women 20--29&9.8&0.8&7.9&12.2&0.4&0.7\\
Average years of schooling of women 30--49&9.1&1.0&6.9&12.1&0.5&0.9\\
Employment rate of women 15--19&16.2&6.9&0.0&50.0&2.7&6.4\\
Employment rate of women 20--29&51.7&8.5&29.3&73.1&5.3&6.8\\
Employment rate of women 30--49&64.4&9.5&38.1&84.8&3.1&9.0\\
Gender pay gap&0.9&0.1&0.4&1.5&0.0&0.1\\
Average disposable income per capita&7180.0&1933.7&3853.4&13994.7&1246.9&1504.7\\
Infant mortality rate&14.7&7.5&0.0&46.4&1.7&7.3\\
\bottomrule
\insertTableNotes
\end{tabularx}
\end{ThreePartTable}
\end{table}
\end{singlespace}
\vspace*{\fill}
\end{landscape}

\begin{figure}[!ht]
\footnotesize
\caption{Evolution of age-specific fertility rates by department in Uruguay (births per 1,000 women, 1976--2021)}
\centering 
\includegraphics[width=0.95\textwidth]{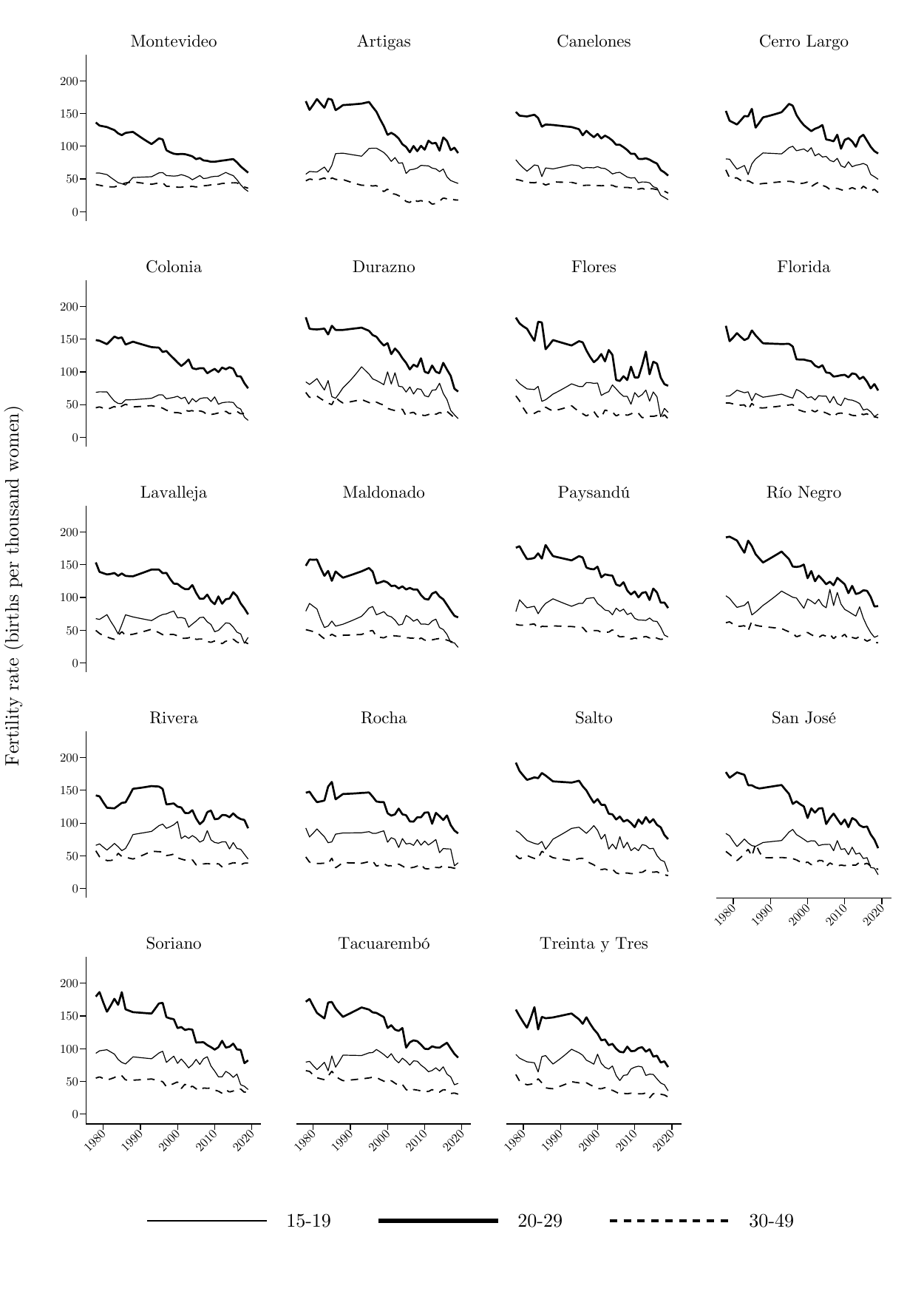}\\
\justifying
\noindent\textit{Source:} Authors' analysis from \textcite{msp2023} and \textcite{wb2023}. 
\label{Figure 3}
\end{figure}

Figure~\ref{Figure 3} illustrates how the age-specific fertility rate fell across the whole national territory over the analysed period. Nevertheless, it also shows that the timing of this decline varies from one department to another.

The number of departments (19) is well below 50, which prevents us from using clustered standard errors to deal with serial correlation issues \parencite{angrist2008}. We therefore follow the advice of \textcite{bekes2021} on dealing with few cross-sectional units in panel data: we employ Newey--West standard errors \parencite{newey1987} that are robust to heteroscedasticity and autocorrelation up to the order suggested by the literature in our left-hand-side variable. In this respect, previous research emphasises that the error term of econometric models analysing the determinants of annual fertility rates tend to follow an AR(1) process \parentext{see, e.g., \textcite{brehm2015} and \textcite{prskawetz2009}}. As a robustness check, we compute the standard errors following the procedure described by \textcite{driscoll1998}, which additionally allows for cross-sectional dependence between departments.

We consider all the variables in levels, without using the log transformation. In this setup, we do not have to worry about stationarity or normality (given the sample size). Furthermore, some departments exhibit zero values for certain variables.

Therefore, we estimate models of the following form:
\begin{equation}
\label{eq 4}
y_{it} = \kappa + \lambda' Z_{it} + \eta_i + \tau_t + \upsilon_{it}
\end{equation}
where $\kappa$ is an intercept, $y_{it}$ is the age-specific fertility rate in logs of department $i$ in year $t$, $Z_{it}$ is a vector containing the time-varying covariates of interest, $\eta_i$ is a department fixed effect, $\tau_t$ is a year fixed effect and $\upsilon_{it}$ represents a time-varying disturbance. It is possible to group the 19 departments into six major geographical regions (metropolitan region, centre, east, northeast, littoral south and littoral north). Our specification allows us to include group-specific linear time trends. This provides a useful robustness check: we can assess whether the results simply follow pre-existing regional trajectories over time. 

\FloatBarrier
\subsection{Results}\label{Section 4.2}

Table~\ref{Table 5} shows the results of our fixed effects panel data analysis. The first column of the table presents the estimates of the model that includes both time and department fixed effects. The second one also includes region-specific linear time trends to assess the robustness of the results. 

According to the estimation results, the share of women in each age group does not have a statistically significant effect on fertility, except that of women aged 15--19. In this case, a one-percentage-point increase in the share of teenagers among women of fertile age raises adolescent fertility by more than four per thousand points. In the context of a declining proportion of adolescents and fertility rates in this age group, a negative and significant coefficient would suggest that the reduction in adolescent fertility is more significant than what would be expected based on population aging alone. This finding may indicate that additional factors beyond demographic shifts, such as changes in social norms or increased access to contraception, are contributing to the decrease in adolescent fertility.

Age-specific nuptiality exerts a statistically significant positive impact on the fertility of women aged 20--29 and especially those aged 15--19. A one-percentage-point increase in the share of women who are married or in a union implies a rise in fertility of almost one point per thousand points for teenagers and about half a point for women aged 25--29. For women aged 30 years old and over, the model with regional time trends shows a statistically significant negative effect of this segment's share on their fertility. Although this finding is remarkable, this variable may capture couples' preferences regarding parenthood (e.g., delaying it for personal or professional reasons). In addition, individuals who marry at a later age are more likely to use contraceptive methods for family planning or health reasons.

The analysis also suggests that the average number of years of schooling leads to a statistically significant reduction in teenage fertility. In the remaining cases, the impact of this variable either is not statistically different from zero or is sensitive to the inclusion of regional trends. 

Regarding the female employment rate, our results are consistent with those presented in the previous section. This variable has a negative significant effect on the fertility only of the oldest group of women. This pattern could have to do with the higher opportunity costs of having children and work--life balance problems for this demographic segment relative to the other, younger ones.

We employ the gender pay gap as an attempt to capture the relationship between women's reproductive and labour market behaviour and decisions. The lack of robustness of these results to the inclusion of regional linear time trends does not allow us to draw any relevant conclusions. Household income appears to have a statistically significant negative impact on the fertility of all segments of women. Several factors may explain this result: higher child-rearing expenses as income rises (e.g., a greater use of private education), larger opportunity costs or even cultural beliefs about parenthood that differ by socioeconomic level (e.g., better-off couples may decide to have a smaller number of children to gain more autonomy in their lives). A 1,000 NCU increase in household income raises the fertility of women aged 15--19 and 30 and over by one point per thousand. The effect is twice as large for women aged 20--29. 

Last, infant mortality has a statistically significant effect on the fertility of women aged 20--29. The impact is null for adolescents and sensitive to the inclusion of regional time trends for the oldest segment of females. In societies with high mortality rates, in the early stages of the demographic transition, one would expect a fall in infant mortality to precede the decline in fertility. This is not the case in Uruguay, where we hypothesise that infant mortality may capture department-level differences in quality of life and access to health care. Families in the territories with the best conditions on these dimensions might show an increased willingness to have children because they perceive better future life chances for their offspring.

The results of the model using Driscoll--Kraay standard errors are remarkably similar to those of our main specification (Table~\ref{Table A9}). 

\begin{landscape}
\begin{singlespace}
\begin{table}[!ht]
\begin{ThreePartTable}
\def\sym#1{\ifmmode^{#1}\else\(^{#1}\)\fi}
\footnotesize
\begin{TableNotes}[flushleft]\setlength\labelsep{0pt}\footnotesize\justifying
\item\textit{Notes}: Standard errors robust to heteroscedasticity and first-order autocorrelation in parentheses. All the models include an intercept. \sym{***} significant at 1\%; \sym{**} significant at 5\%; \sym{*} significant at 10\%.
\end{TableNotes}
\begin{tabularx}{\linewidth}{X *{6}{@{}S[table-column-width=2.1cm]}}
\caption{Estimation results of the panel data model for regional fertility rates} \label{Table 5}\\
\toprule
&\multicolumn{1}{c}{(I)}&\multicolumn{1}{c}{(II)}&\multicolumn{1}{c}{(III)}&\multicolumn{1}{c}{(IV)}&\multicolumn{1}{c}{(V)}&\multicolumn{1}{c}{(VI)}\\[1ex]
&\multicolumn{6}{c}{Fertility rate of}\\[1ex]
&\multicolumn{2}{c}{women aged 15--19}&\multicolumn{2}{c}{women aged 20--29}&\multicolumn{2}{c}{women aged 30 and above}\\
\midrule
\% of women in the age bracket&-4.336\sym{***}&-4.253\sym{***}&-0.410&-0.023&-0.012&0.045\\
&(1.036)&(0.909)&(0.884)&(0.798)&(0.457)&(0.347)\\ 
\% of women married (age-specific)&0.852\sym{*}&0.978\sym{**}&0.445\sym{***}&0.570\sym{***}&-0.086&-0.132\sym{**}\\
&(0.446)&(0.431)&(0.166)&(0.170)&(0.067)&(0.060)\\ 
Average years of schooling&-4.341\sym{***}&-3.445\sym{**}&0.716&0.500&4.390\sym{***}&0.868\\
&(1.480)&(1.564)&(1.042)&(0.994)&(0.681)&(0.347)\\ 
Age-specific female employment rate&0.108&0.077&0.099&0.017&-0.327\sym{***}&-0.150\sym{***}\\
&(0.066)&(0.065)&(0.074)&(0.068)&(0.064)&(0.049)\\ 
Gender pay gap&9.676\sym{***}&4.113&0.839&-3.580&4.356\sym{*}&1.088\\
&(3.379)&(2.729)&(4.152)&(3.579)&(2.283)&(1.816)\\ 
Average disposable income per capita&-0.001&-0.001\sym{**}&-0.002\sym{***}&-0.002\sym{***}&-0.001\sym{***}&-0.001\sym{***}\\
&(0.001)&(0.000)&(0.001)&(0.001)&(0.000)&(0.000)\\ 
Infant mortality rate&-0.201\sym{*}&-0.159&-0.275\sym{**}&-0.244\sym{***}&-0.083&-0.112\sym{**}\\
&(0.110)&(0.102)&(0.110)&(0.097)&(0.062)&(0.053)\\[1ex]
Year fixed effects&\multicolumn{1}{c}{\checkmark}&\multicolumn{1}{c}{\checkmark}&\multicolumn{1}{c}{\checkmark}&\multicolumn{1}{c}{\checkmark}&\multicolumn{1}{c}{\checkmark}&\multicolumn{1}{c}{\checkmark}\\
Department fixed effects&\multicolumn{1}{c}{\checkmark}&\multicolumn{1}{c}{\checkmark}&\multicolumn{1}{c}{\checkmark}&\multicolumn{1}{c}{\checkmark}&\multicolumn{1}{c}{\checkmark}&\multicolumn{1}{c}{\checkmark}\\
Region-specific linear time trends&&\multicolumn{1}{c}{\checkmark}&&\multicolumn{1}{c}{\checkmark}&&\multicolumn{1}{c}{\checkmark}\\[1ex]
$R^2$&0.866&0.885&0.943&0.951&0.678&0.799\\
No. of observations&\multicolumn{1}{c}{\hspace{6.5mm}684}&\multicolumn{1}{c}{\hspace{6.5mm}684}&\multicolumn{1}{c}{\hspace{6.5mm}684}&\multicolumn{1}{c}{\hspace{6.5mm}684}&\multicolumn{1}{c}{\hspace{6.5mm}684}&\multicolumn{1}{c}{\hspace{4mm}684}\\
Mean of dependent variable&61.718&61.718&107.679&107.679&41.283&41.283\\
\bottomrule 
\insertTableNotes
\end{tabularx}
\end{ThreePartTable}
\end{table}
\end{singlespace}
\end{landscape}

\FloatBarrier
\section{Conclusion}\label{Section 5} 

This study has examined a range of factors that influence fertility throughout the reproductive cycle, using time series data from 1968 to 2021 and panel data from regional statistical sources from 1984 to 2019. We have focused on fertility across three stages: adolescent (15 to 19 years), intermediate fertility (20 to 29 years) and late fertility (30 to 49 years).

The analysis conducted in these pages has allowed testing some of the theoretical hypotheses put forward in the literature. Although the nature of this study is eminently descriptive, the use of multiple methods and remarkably long data series enhances our understanding of fertility behaviour.

Compared to most Latin American and Caribbean countries, Uruguay has atypical characteristics. It underwent the first demographic transition at an early stage. For decades, it has had quite low fertility rates among women aged 20 and over but, until recently, experienced a relatively high teenage birth rate, especially among households of low socioeconomic status. Moreover, most births take place outside of marriage, against a background of rising divorce rates. These features have led some observers to suggest that the country is undergoing a second demographic transition.

This work has considered a range of socioeconomic indicators---GDP per capita and department-level average household income per capita---to test the relevance of some of the conventional structural or diffusion, maternal role incompatibility and institutional theories. Specifically, the study has used two economic indicators that reflect the social and economic conditions that shape the lives of Uruguayan women. On the one hand, GDP per capita is a better approximation of economic cycles, although the literature suggests the existence of time lags and differences throughout a woman's reproductive cycle. On the other hand, departmental average per capita income better captures the actual appropriation of economic output by households.

Previous literature suggests that the relationship between the level of income or GDP per capita and fertility is complex. In the long term, one expects a negative association, but short-run shocks might shape this relationship, which may also vary by age group. In the time series analysis, we have detected such differences between short- and long-run impacts only for the fertility of women aged 30 years old and above. The results of the panel data econometric exercise (which does not allow us to distinguish between short- and long-term effects) indicate a negative association between income and fertility at all ages. These estimates may well capture greater availability of and access to contraceptive methods, greater educational opportunities and an increase in the opportunity cost of having children.

Another relevant dimension that allows us to understand reproductive decisions is female employment because of the potential conflict between professional career and motherhood. The increase in women's employment may have a negative impact on fertility, as women might decide to postpone motherhood or reduce the number of children to prioritise their labour market performance. Contrary to expectations, our time series analysis has shown a negative effect of female employment on fertility. Nevertheless, the econometric exercise using regional data, which allows us to control for unobserved heterogeneity and age-specific female employment rates, has revealed a negative impact. This result could reflect the opportunity costs of children, which raises the relevance of designing social protection policies that alleviate work--life balance problems, such as making affordable childcare widely available or providing appropriate parental leave.

The relationship between education and its impact on fertility has received extensive attention in the specialised literature. Overall, existing studies suggest a negative association between schooling levels and fertility for a variety of reasons, ranging from greater access to information, better job opportunities or changes in culture, social norms or preferences for motherhood. Our results have confirmed this relationship, particularly in the case of adolescent fertility. The possibility of curbing teenage motherhood by expanding education has emerged as a clear policy implication of this analysis.

Given Uruguay's stage of the demographic transition, as argued above, one should expect a positive or nonsignificant relationship between fertility and infant mortality. However, our results for this variable are not robust across estimation methods and analysed periods. This lack of conclusiveness in our findings could indicate that this variable reflects characteristics related to economic conditions and health care that are not captured by other covariates.

Finally, the sizeable fall in births in recent years---especially among teenagers---might also have been the consequence of different national policies implemented since 2008. Such government initiatives include the 2008 health care reform (which moved health care towards an integrated system), a new law on sexual and reproductive health in 2008, the setting of health care targets related to teenagers in 2010, the expansion of contraceptive methods fully or heavily subsidised by the health care system in 2011, the decriminalisation of abortion in 2012, initiatives to promote youth participation in civic life in 2014 or the rollout of a new strategy to prevent teenage pregnancy in 2016. Whereas these policies have received a great deal of attention in various studies demonstrating their relevance \parentext{see, e.g., \textcite{anton2018}, \textcite{cabella2022}, \textcite{ceni2021} or \textcite{balsa2021}}, our research design, constrained by the number observations and the contemporaneous nature of the mentioned interventions, cannot adequately include them in the analysis and disentangle their causal effects. Therefore, they should be the object of subsequent separate research works.

\clearpage
\singlespacing
\printbibliography

\clearpage
\appendix
\section*{Supplementary appendix}\label{Appendix}
\setcounter{page}{1}
\renewcommand\thepage{S\arabic{page}}
\setcounter{table}{0}
\setcounter{figure}{0}
\renewcommand\thetable{A.\arabic{table}}
\renewcommand\thefigure{A.\arabic{figure}} 

\vspace*{\fill}
\begin{singlespace}
\begin{table}[!ht]
\begin{ThreePartTable}
\def\sym#1{\ifmmode^{#1}\else\(^{#1}\)\fi}
\scriptsize
\begin{TableNotes}[flushleft]\setlength\labelsep{0pt}\footnotesize\justifying
\item\textit{Notes}: The results correspond to models with a constant and without a linear time trend. They remain the same when a linear time trend is included.
\end{TableNotes}
\begin{tabularx}{\textwidth}{X *{8}{S[table-column-width=1cm]}}
\caption{Results of unit-root tests} \label{Table A1}\\
\toprule
&\multicolumn{4}{c}{In levels}&\multicolumn{4}{c}{In first differences}\\ [1ex]
&\multicolumn{2}{c}{ADF test}&\multicolumn{2}{c}{PP test}&\multicolumn{2}{c}{ADF test}&\multicolumn{2}{c}{PP test}\\ [1ex] 
&\multicolumn{1}{c}{\textit{t}}&\multicolumn{1}{c}{\textit{p}-value}&\multicolumn{1}{c}{\textit{t}}&\multicolumn{1}{c}{\textit{p}-value}&\multicolumn{1}{c}{\textit{t}}&\multicolumn{1}{c}{\textit{p}-value}&\multicolumn{1}{c}{\textit{t}}&\multicolumn{1}{c}{\textit{p}-value}\\ 
\midrule
$\log \left(\text{Fertility 15--19}\right)$&5.475&1.000&3.250&1.000&-2.871&0.049&-2.856&0.051\\ 
$\log \left(\text{Fertility 20--29}\right)$&2.333&0.999&1.656&0.998&-3.680&0.004&-3.700&0.004\\
$\log \left(\text{Fertility 30+}\right)$&0.412&0.982&-0.032&0.956&-4.647&0.000&-4.630&0.000\\ 
$\log \left(\text{GDP per capita}\right)$&0.424&0.982&0.139&0.969&-5.109&0.000&-5.043&0.000\\ 
$\log \left(\text{employment rate}\right)$&-1.639&0.463&-1.639&0.463&-6.962&0.000&-6.969&0.000\\ 
$\log \left(\text{enrolment rate}\right)$&-0.089&0.951&-0.208&0.938&-5.777&0.000&-5.729&0.000\\
$\log \left(\text{Infant mortality}\right)$&0.422&0.982&0.836&0.992&-9.662&0.000&-9.907&0.000\\ 
\bottomrule
\insertTableNotes 
\end{tabularx}
\end{ThreePartTable}
\end{table}
\end{singlespace}
\vspace*{\fill}
\clearpage

\vspace*{\fill}
\begin{singlespace}
\begin{table}[!ht]
\begin{ThreePartTable}
\def\sym#1{\ifmmode^{#1}\else\(^{#1}\)\fi}
\footnotesize
\setlength\tabcolsep{0.5em}
\begin{TableNotes}[flushleft]\setlength\labelsep{0pt}\footnotesize\justifying
\item\textit{Note}: If rank = 0, there is no cointegration relationship; if rank = 1, there is at least one cointegration relationship, and so on.
\end{TableNotes}
\begin{tabularx}{\textwidth}{X *{6}{S[table-column-width=0.1cm]}}
\caption{Results of Johansen test for cointegration} \label{Table A2}\\
\toprule
&\multicolumn{6}{c}{Fertility rate of women aged}\\[1ex]
&\multicolumn{2}{c}{15--19}&\multicolumn{2}{c}{20--29}&\multicolumn{2}{c}{30 and above}\\[1ex]
&\multicolumn{1}{c}{Statistic}&\multicolumn{1}{c}{\makecell{5\% critical\\value}}&\multicolumn{1}{c}{Statistic}&\multicolumn{1}{c}{\makecell{5\% critical\\value}}&\multicolumn{1}{c}{Statistic}&\multicolumn{1}{c}{\makecell{5\% critical\\value}}\\
\midrule
Trace&&&&&&\\
~~Rank = 0&72.235&68.52&79.843&68.52&80.780&68.52\\
~~Rank = 1&41.241&47.21&45.862&47.21&50.889&47.21\\
~~Rank = 2&19.058&29.68&25.459&29.68&27.221&29.68\\
~~Rank = 3&5.535&15.41&10.221&15.41&11.291&15.41\\
~~Rank = 4&0.106&3.76&0.012&3.76&0.309&3.76\\ [1ex] 
Maximum eigenvalue&&&&&&\\
~~Rank = 0&30.993&33.46&33.982&33.46&29.891&33.46\\
~~Rank = 1&22.183&27.07&20.403&27.07&23.668&27.07\\
~~Rank = 2&13.523&20.97&15.238&20.97&15.930&20.97\\
~~Rank = 3&5.429&14.07&10.209&14.07&10.983&14.07\\
~~Rank = 4&0.106&3.76&0.012&3.76&0.309&3.76\\ 
\bottomrule
\insertTableNotes 
\end{tabularx}
\end{ThreePartTable}
\end{table}
\end{singlespace}
\vspace*{\fill}
\clearpage
\vspace*{\fill}
\begin{singlespace}
\begin{table}[!ht]
\begin{ThreePartTable}
\def\sym#1{\ifmmode^{#1}\else\(^{#1}\)\fi}
\footnotesize
\begin{tabularx}{\textwidth}{X *{6}{S[table-column-width=1cm]}}
\caption{Results of the Engle and Granger test for cointegration} \label{Table A3}\\
\toprule
&\multicolumn{6}{c}{Fertility rate of women aged}\\[1ex]
&\multicolumn{2}{c}{15--19}&\multicolumn{2}{c}{20--29}&\multicolumn{2}{c}{30 and above}\\[1ex]
&\multicolumn{1}{c}{\textit{t}-statistic}&\multicolumn{1}{c}{\textit{p}-value}&\multicolumn{1}{c}{\textit{t}-statistic}&\multicolumn{1}{c}{\textit{p}-value}&\multicolumn{1}{c}{\textit{t}-statistic}&\multicolumn{1}{c}{\textit{p}-value}\\
\midrule
ADF test&-3.841&0.015&-3.941&0.011&-4.836&0.000\\
PP test&-3.867&0.014&-3.936&0.011&-4.830&0.000\\
\bottomrule
\end{tabularx}
\end{ThreePartTable}
\end{table}
\end{singlespace}
\vspace*{\fill}
\clearpage

\vspace*{\fill}
\begin{figure}[!ht]
\footnotesize
\caption{Cumulative sum of squares tests for parameter stability}
\centering 
\includegraphics[width=1\textwidth]{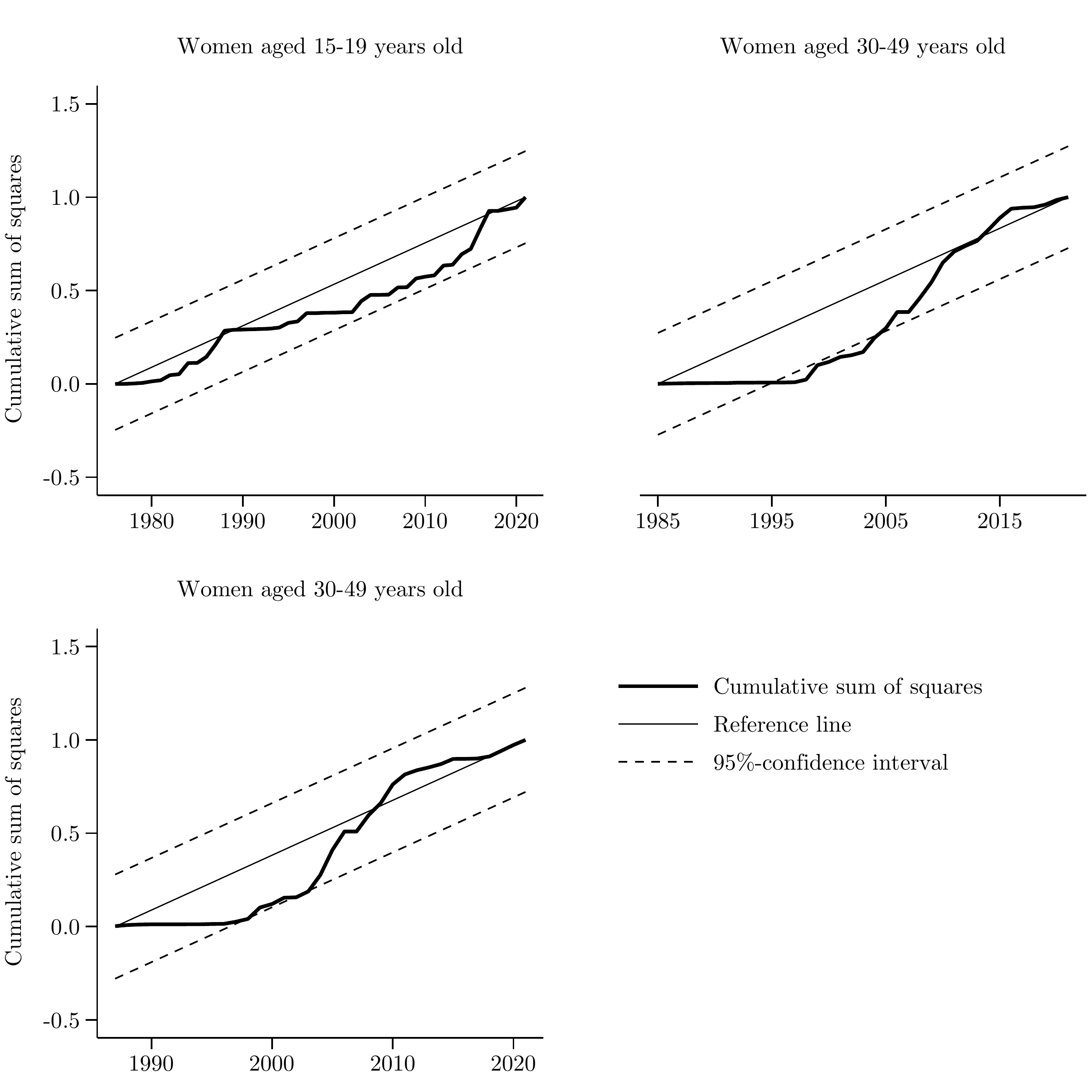}\\
\justifying
\label{Figure A1}
\end{figure}
\vspace*{\fill}
\clearpage

\vspace*{\fill}
\begin{singlespace}
\begin{table}[!ht]
\begin{ThreePartTable}
\def\sym#1{\ifmmode^{#1}\else\(^{#1}\)\fi}
\footnotesize
\begin{tabularx}{0.85\textwidth}{X *{3}{S[table-column-width=2cm]}}
\caption{Results of Pesaran, Shin and Smith test for cointegration} \label{Table A4}\\
\toprule
&\multicolumn{3}{c}{Fertility rate of women aged}\\[1ex]
&\multicolumn{1}{c}{15--19}&\multicolumn{1}{c}{20--29}&\multicolumn{1}{c}{30 and above}\\
\midrule
\textit{F}-statistic&6.242&3.138&3.968\\
~~\textit{p}-value $I(0)$&0.003&0.113&0.001\\
~~\textit{p}-value $I(1)$&0.016&0.310&0.005\\[1ex] 
\textit{t}-statistic&4.452&3.043&1.594\\
~~\textit{p}-value $I(0)$&0.004&0.106&0.307\\
~~\textit{p}-value $I(1)$&0.045&0.353&0.606\\[1ex] 
Residual degrees of freedom&\multicolumn{1}{c}{\hspace{1.5mm}41}&\multicolumn{1}{c}{\hspace{1.5mm}31}&\multicolumn{1}{c}{\hspace{1.5mm}29}\\ 
Model degrees of freedom&\multicolumn{1}{c}{\hspace{1.5mm}10}&\multicolumn{1}{c}{\hspace{1.5mm}10}&\multicolumn{1}{c}{\hspace{1.5mm}12}\\
No. of observations&\multicolumn{1}{c}{\hspace{1.5mm}52}&\multicolumn{1}{c}{\hspace{1.5mm}42}&\multicolumn{1}{c}{\hspace{1.5mm}42}\\ 
ARDL model structure&\multicolumn{1}{l}{\hspace{1.5mm}(2,0,0,0,2)}&\multicolumn{1}{l}{\hspace{1.5mm}(2,1,0,0,1)}&\multicolumn{1}{l}{\hspace{1.5mm}(1,0,0,0,1)}\\ 
\bottomrule
\end{tabularx}
\end{ThreePartTable}
\end{table}
\end{singlespace}
\vspace*{\fill}
\clearpage

\begin{landscape}
\vspace*{\fill}
\begin{singlespace}
\begin{table}[!ht]
\begin{ThreePartTable}
\def\sym#1{\ifmmode^{#1}\else\(^{#1}\)\fi}
\footnotesize
\begin{TableNotes}[flushleft]\setlength\labelsep{0pt}\footnotesize\justifying
\item\textit{Note}: The structures of the three models in the table are ARDL(2,0,0,0,2), ARDL(2,1,0,0,1) and ARDL(1,2,0,0,1), respectively.
\end{TableNotes}
\begin{tabularx}{\linewidth}{X *{6}{S[table-column-width=1.75cm]}}
\caption{Results of tests of goodness of fit} \label{Table A5}\\
\toprule
&\multicolumn{6}{c}{Fertility rate of women aged}\\[1ex]
&\multicolumn{2}{c}{15--19}&\multicolumn{2}{c}{20--29}&\multicolumn{2}{c}{30 and above}\\[1ex]
&\multicolumn{1}{c}{Statistic}&\multicolumn{1}{c}{\textit{p}-value}&\multicolumn{1}{c}{Statistic}&\multicolumn{1}{c}{\textit{p}-value}&\multicolumn{1}{c}{Statistic}&\multicolumn{1}{c}{\textit{p}-value}\\
\midrule
\makecell[tl]{First-order autocorrelation\\test (Breusch--Godfrey)}&2.163&0.149&0.410&0.527&1.570&0.220\\
\makecell[tl]{First-order autocorrelation\\test (Durbin--Watson)}&1.736&0.195&0.295&0.591&1.099&0.303\\ 
\makecell[tl]{Heteroscedasticity test\\(White)}&52.000&0.435&42.000&0.427&43.000&0.428\\
\makecell[tl]{Heteroscedasticity test\\(Breusch--Pagan/Cook--Weisberg)}&0.934&0.334&0.511&0.475&0.034&0.854\\
\makecell[tl]{Normality test\\
(D'Agostino et al.)}&0.750&0.703&2.773&0.250&0.033&0.984\\
\bottomrule
\insertTableNotes
\end{tabularx} 
\end{ThreePartTable}
\end{table}
\end{singlespace}
\vspace*{\fill}
\end{landscape}
\clearpage

\vspace*{\fill}
\begin{singlespace}
\begin{table}[!ht]
\centering
\begin{threeparttable}
\def\sym#1{\ifmmode^{#1}\else\(^{#1}\)\fi}
\footnotesize
\caption{Variance decomposition of VAR model for fertility rate of women between 15 and 19 years old}
\label{Table A6} 
\begin{tabular}{S *{5}{S[table-column-width=1.9cm]}} 
\toprule
\multicolumn{1}{c}{\makecell{Time \\period}}&\multicolumn{1}{c}{\makecell{log (Fertility\\rate\\15--19)}}&\multicolumn{1}{c}{\makecell{log (GDP\\per capita)}}&\multicolumn{1}{c}{\makecell{log (Female\\employment\\rate)}}&\multicolumn{1}{c}{\makecell{log (Female\\ high school\\enrolment)}}&\multicolumn{1}{c}{\makecell{log (Infant\\mortality\\rate)}}\\
\midrule
1&1.000&0.000&0.000&0.000&0.000\\
&(0.000)&(0.000)&(0.000)&(0.000)&(0.000)\\ 
2&0.766&0.050&0.031&0.108&0.044\\
&(0.090)&(0.051)&(0.035)&(0.060)&(0.046)\\ 
3&0.751&0.083&0.052&0.084&0.030\\
&(0.103)&(0.073)&(0.051)&(0.056)&(0.030)\\
4&0.760&0.082&0.052&0.079&0.027\\ 
&(0.115)&(0.082)&(0.055)&(0.059)&(0.029)\\
5&0.764&0.078&0.055&0.074&0.029\\
&(0.115)&(0.076)&(0.060)&(0.057)&(0.025)\\
6&0.767&0.077&0.055&0.073&0.029\\
&(0.115)&(0.073)&(0.061)&(0.056)&(0.024)\\
7&0.767&0.076&0.055&0.072&0.030\\ 
&(0.116)&(0.073)&(0.062)&(0.056)&(0.024)\\
8&0.768&0.076&0.055&0.071&0.030\\ 
&(0.117)&(0.073)&(0.062)&(0.056)&(0.024)\\
\bottomrule
\end{tabular}
\begin{tablenotes}[flushleft]
\setlength\labelsep{0pt}
\footnotesize 
\item\textit{Note}: Standard errors in parentheses.
\end{tablenotes}
\end{threeparttable}
\end{table}
\end{singlespace}
\vspace*{\fill}
\clearpage

\vspace*{\fill}
\begin{singlespace}
\begin{table}[!ht]
\centering
\begin{threeparttable}
\def\sym#1{\ifmmode^{#1}\else\(^{#1}\)\fi}
\footnotesize
\caption{Variance decomposition of VAR model for fertility rate of women between 20 and 29 years old} 
\label{Table A7} 
\begin{tabular}{S *{5}{S[table-column-width=1.9cm]}} 
\toprule
\multicolumn{1}{c}{\makecell{Time \\period}}&\multicolumn{1}{c}{\makecell{log (Fertility\\rate\\20--29)}}&\multicolumn{1}{c}{\makecell{log (GDP\\per capita)}}&\multicolumn{1}{c}{\makecell{log (Female\\employment\\rate)}}&\multicolumn{1}{c}{\makecell{log (Female\\ high school\\enrolment)}}&\multicolumn{1}{c}{\makecell{log (Infant\\mortality\\rate)}}\\
\midrule
1&1.000&0.000&0.000 &0.000&0.000\\ 
&(0.000)&(0.000)&(0.000)&(0.000)&(0.000)\\
2&0.955&0.001&0.012&0.024&0.008\\ 
&(0.036)&(0.007)&(0.016)&(0.025)&(0.015)\\
3&0.929&0.005&0.013&0.049&0.005\\ 
&(0.062)&(0.013)&(0.023)&(0.050)&(0.011)\\
4&0.901&0.017&0.012&0.064&0.006\\ 
&(0.088)&(0.036)&(0.026)&(0.067)&(0.016)\\
5&0.876&0.032&0.011&0.073&0.008\\ 
&(0.109)&(0.057)&(0.028)&(0.077)&(0.020)\\
6&0.860&0.041x&0.011&0.079&0.\\ 
&(0.122)&(0.070)&(0.029)&(0.083)&(0.023)\\
7&0.850&0.046&0.010&0.083&0.010\\ 
&(0.132)&(0.078)&(0.030)&(0.089)&(0.025)\\
8&0.843&0.050&0.010&0.087&0.010\\ 
&(0.140)&(0.084)&(0.031)&(0.093)&(0.026)\\
\bottomrule
\end{tabular}
\begin{tablenotes}[flushleft]
\setlength\labelsep{0pt}
\footnotesize 
\item\textit{Note}: Standard errors in parentheses.
\end{tablenotes}
\end{threeparttable}
\end{table}
\end{singlespace}
\vspace*{\fill}
\clearpage

\vspace*{\fill}
\begin{singlespace}
\begin{table}[!ht]
\centering
\begin{threeparttable}
\def\sym#1{\ifmmode^{#1}\else\(^{#1}\)\fi}
\footnotesize
\caption{Variance decomposition of VAR model for fertility rate of women aged 30 years old or more} 
\label{Table A8} 
\begin{tabular}{S *{5}{S[table-column-width=1.9cm]}} 
\toprule
\multicolumn{1}{c}{\makecell{Time \\period}}&\multicolumn{1}{c}{\makecell{log (Fertility\\rate\\20--29)}}&\multicolumn{1}{c}{\makecell{log (GDP\\per capita)}}&\multicolumn{1}{c}{\makecell{log (Female\\employment\\rate)}}&\multicolumn{1}{c}{\makecell{log (Female\\ high school\\enrolment)}}&\multicolumn{1}{c}{\makecell{log (Infant\\mortality\\rate)}}\\
\midrule
1&1.000&0.000&0.000&0.000&0.000\\
&(0.000)&(0.000)&(0.000)&(0.000)&(0.000)\\
2&0.962&0.010&0.019 &0.004&0.006\\
&(0.033)&(0.018)&(0.021)&(0.010)&(0.013)\\
3&0.920&0.018&0.027&0.032&0.004\\
&(0.068)&(0.035)&(0.036)&(0.041)&(0.011)\\
4&0.881&0.021&0.036&0.057&0.004\\
&(0.100)&(0.046)&(0.051)&(0.064)&(0.013)\\
5&0.862&0.020&0.042&0.070&0.007\\
&(0.119)&(0.048)&(0.062)&(0.077)&(0.020)\\
6&0.853&0.018&0.043&0.078&0.007\\
&(0.129)&(0.049)&(0.067)&(0.084)&(0.023)\\
7&0.847&0.019&0.044&0.082&0.008\\
&(0.135)&(0.053)&(0.070)&(0.089)&(0.024)\\
8&0.842&0.019&0.045&0.085&0.008\\
&(0.141)&(0.056)&(0.073)&(0.092)&(0.026)\\
\bottomrule
\end{tabular}
\begin{tablenotes}[flushleft]
\setlength\labelsep{0pt}
\footnotesize 
\item\textit{Note}: Standard errors in parentheses.
\end{tablenotes}
\end{threeparttable}
\end{table}
\end{singlespace}
\vspace*{\fill}

\vspace*{\fill}
\begin{figure}[!ht]
\centering
\begin{minipage}[c]{0.7\textwidth}
\centering
\footnotesize
\caption{Impulse response functions to a one-standard-deviation shock to covariates and fertility itself}
\centering 
\includegraphics[width=1\textwidth]{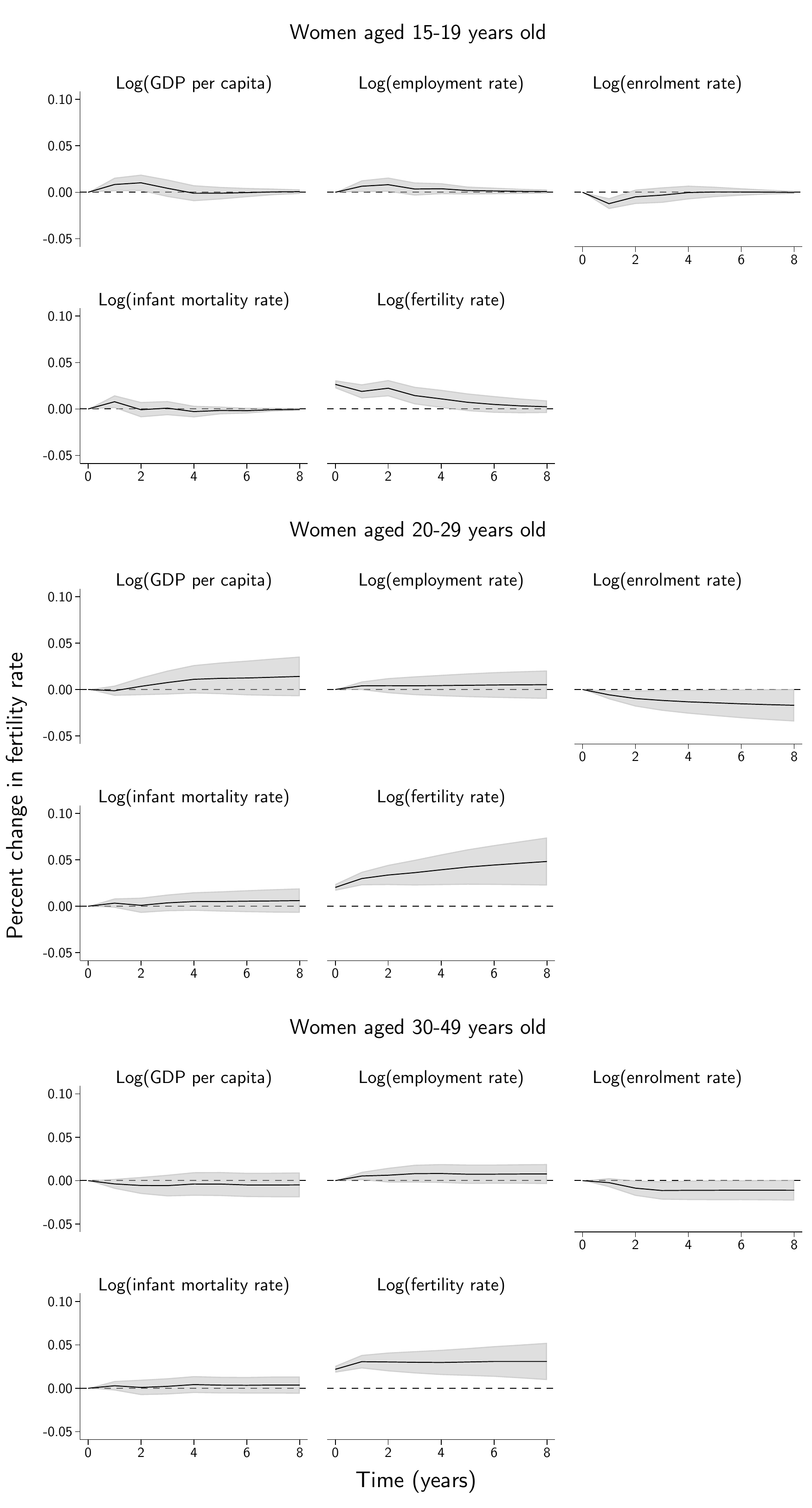}\\
\justifying
\noindent\textit{Note:} The grey-shaded areas indicate 90\% confidence intervals. 
\label{Figure A2}
\end{minipage}
\end{figure}
\vspace*{\fill}
\clearpage

\begin{landscape}
\begin{singlespace}
\begin{table}[!ht]
\begin{ThreePartTable}
\def\sym#1{\ifmmode^{#1}\else\(^{#1}\)\fi}
\footnotesize
\begin{TableNotes}[flushleft]\setlength\labelsep{0pt}\footnotesize\justifying
\item\textit{Notes}: Standard errors robust to heteroscedasticity, first-order autocorrelation and cross-sectional dependence in parentheses. All the models include an intercept. \sym{***} significant at 1\%; \sym{**} significant at 5\%; \sym{*} significant at 10\%.
\end{TableNotes}
\begin{tabularx}{\linewidth}{X *{6}{@{}S[table-column-width=2.1cm]}}
\caption{Estimation results of panel data model for regional fertility rates with Driscoll--Kraay standard errors} \label{Table A9}\\
\toprule
&\multicolumn{1}{c}{(I)}&\multicolumn{1}{c}{(II)}&\multicolumn{1}{c}{(III)}&\multicolumn{1}{c}{(IV)}&\multicolumn{1}{c}{(V)}&\multicolumn{1}{c}{(VI)}\\[1ex]
&\multicolumn{6}{c}{Fertility rate of}\\[1ex]
&\multicolumn{2}{c}{women aged 15--19}&\multicolumn{2}{c}{women aged 20--29}&\multicolumn{2}{c}{women aged 30 and above}\\
\midrule
\% of women in the age bracket&-4.336\sym{***}&-4.253\sym{***}&-0.410&-0.023&-0.012&0.045\\
&(1.421)&(1.014)&(0.990)&(0.698)&(0.539)&(0.427)\\ 
\% of women married (age-specific)&0.851\sym{*}&0.978\sym{**}&0.445\sym{***}&0.570\sym{***}&-0.086&-0.132\sym{**}\\
&(0.422)&(0.363)&(0.131)&(0.115)&(0.060)&(0.046)\\ 
Average years of schooling&-4.341\sym{**}&-3.445\sym{**}&0.716&0.500&4.390\sym{***}&0.868\\
&(1.556)&(1.574)&(1.096)&(1.108)&(0.911)&(0.678)\\ 
Age-specific female employment rate&0.108\sym{*}&0.077&0.099\sym{*}&0.017\sym{*}&-0.327\sym{***}&-0.150\sym{**}\\
&(0.053)&(0.055)&(0.056)&(0.056)&(0.082)&(0.056)\\ 
Gender pay gap&9.676\sym{**}&4.113&0.839&-3.580&4.356\sym{*}&1.088\\
&(3.698)&(2.992)&(2.113)&(4.881)&(3.840)&(1.756)\\ 
Average disposable income per capita&-0.001&-0.001\sym{**}&-0.002\sym{***}&-0.002\sym{***}&-0.001\sym{***}&-0.001\sym{**}\\
&(0.001)&(0.001)&(0.001)&(0.001)&(0.001)&(0.000)\\ 
Infant mortality rate&-0.201&-0.159&-0.275\sym{*}&-0.244\sym{**}&-0.083&-0.112\sym{**}\\
&(0.140)&(0.123)&(0.053)&(0.138)&(0.105)&(0.047)\\[1ex]
Year fixed effects&\multicolumn{1}{c}{\checkmark}&\multicolumn{1}{c}{\checkmark}&\multicolumn{1}{c}{\checkmark}&\multicolumn{1}{c}{\checkmark}&\multicolumn{1}{c}{\checkmark}&\multicolumn{1}{c}{\checkmark}\\
Department fixed effects&\multicolumn{1}{c}{\checkmark}&\multicolumn{1}{c}{\checkmark}&\multicolumn{1}{c}{\checkmark}&\multicolumn{1}{c}{\checkmark}&\multicolumn{1}{c}{\checkmark}&\multicolumn{1}{c}{\checkmark}\\
Region-specific linear time trends&&\multicolumn{1}{c}{\checkmark}&&\multicolumn{1}{c}{\checkmark}&&\multicolumn{1}{c}{\checkmark}\\[1ex]
$R^2$&0.866&0.885&0.943&0.951&0.678&0.799\\
No. of observations&\multicolumn{1}{c}{\hspace{6.5mm}684}&\multicolumn{1}{c}{\hspace{6.5mm}684}&\multicolumn{1}{c}{\hspace{6.5mm}684}&\multicolumn{1}{c}{\hspace{6.5mm}684}&\multicolumn{1}{c}{\hspace{6.5mm}684}&\multicolumn{1}{c}{\hspace{4mm}684}\\
Mean of dependent variable&61.718&61.718&107.679&107.679&41.283&41.283\\
\bottomrule 
\insertTableNotes
\end{tabularx}
\end{ThreePartTable}
\end{table}
\end{singlespace}
\end{landscape}

\end{document}